\documentclass[prb,twocolumn]{revtex4}%

\usepackage{graphicx}
\usepackage{amsmath}
\usepackage{ulem}
\usepackage{color}

\newcommand{\be}{\begin{equation}}
\newcommand{\ee}{\end{equation}}
\newcommand{\bea}{\begin{eqnarray*}}
\newcommand{\eea}{\end{eqnarray*}}
\newcommand{\bean}{\begin{eqnarray}}
\newcommand{\eean}{\end{eqnarray}}

\begin{document}

\draft
\title{\bf
Topological Interface States and Nonlinear Thermoelectric
Performance in Armchair Graphene Nanoribbon Heterostructures}

\author{David M T Kuo}

\address{Department of Electrical Engineering and Department of Physics, National Central
University, Chungli, 32001 Taiwan}

\date{\today}

\begin{abstract}
We investigate the emergence and topological nature of interface
states (IFs) in N-AGNR/$(N-2)$-AGNR/N-AGNR heterostructure (AGNRH)
segments lacking translational symmetry, focusing on their
relation to the end states (ESs) of the constituent armchair
graphene nanoribbon (AGNR) segments. For AGNRs with $R_1$-type
unit cells, the ES numbers under a longitudinal electric field
follow the relations $N = N_{A(B)} \times 6 + 1$ and $N = N_{A(B)}
\times 6 + 3$, whereas $R_2$-type unit cells exhibit $(N_{A(B)} +
1)$ ESs. The subscripts $A$ and $B$ denote the chirality types of
the ESs. The Stark effect lifts ES degeneracy and enables clear
spectral separation between ESs and IFs. Using a real-space bulk
boundary perturbation approach, we show that opposite-chirality
states hybridize through junction-site perturbations and may shift
out of the bulk gap. The number and chirality of IFs in symmetric
AGNRHs are determined by the difference between the ESs of the
outer and central segments, $N_O$ and $N_C$, according to
$N_{IF,\beta} = |N_{O,B(A)} - N_{C,A(B)}|$, where $\beta$ labels
the chirality. Depending on whether $N_O > N_C$ or $N_C > N_O$,
the resulting IFs acquire B- or A-chirality, respectively.
Calculated transmission spectra ${\cal T}_{GNR}(\varepsilon)$
reveal that AGNRHs host a topological double quantum dot (TDQD)
when IFs originate from the ESs of the central AGNR segment. Using
an Anderson model with effective intra-dot and inter-dot Coulomb
interactions, we derive an analytical expression for the tunneling
current through the TDQD via a closed-form transmission
coefficient. Thermoelectric analysis shows that TDQDs yield
enhanced nonlinear power output in the electron-dilute and
hole-dilute charge states, with Coulomb blockade suppressing
thermal current but not thermal voltage. The thermal power output
of the TDQD is significantly enhanced by nonlinear effects, even
under strong electron Coulomb interactions.
\end{abstract}

\maketitle

\section{Introduction}
Since the groundbreaking discovery of graphene in 2004
[\onlinecite{Novoselovks}], extensive experimental and theoretical
studies have focused on graphene nanoribbons (GNRs)
[\onlinecite{Cai}--\onlinecite{SongST}]. Advances in bottom-up
fabrication now allow atomically precise GNRs with diverse
geometries [\onlinecite{Cai}--\onlinecite{SongST}]. Among these,
armchair GNRs (AGNRs) exhibit width-dependent electronic
structures [\onlinecite{Cai}-\onlinecite{Nestor}], enabling
tunable semiconducting phases suitable for quantum device
applications. Of particular interest are zero-dimensional
topological states (0D-TSs) associated with the interface states
(IFs) of AGNR heterostructures (AGNRHs), which have been observed
in scanning tunneling microscopy measurements of the local
electronic density [\onlinecite{Groning}--\onlinecite{DJRizzo}].
These 0D-TSs emerge within the mid-gap region of semiconducting
AGNRHs [\onlinecite{DRizzo}--\onlinecite{DJRizzo}], providing a
promising route toward atomically precise topological quantum dot
(TQD) devices [\onlinecite{LeobandungE}--\onlinecite{PerrinML}].
Their robustness allows the design of double TQDs and TQD arrays
with controllable electron hopping and Coulomb interactions
between TSs [\onlinecite{DJRizzo}].

Topological states in GNRs were originally predicted using the Zak
number $Z_2$ of bulk GNRs [\onlinecite{CaoT}], which requires
time-reversal and translational symmetries
[\onlinecite{CaoT}--\onlinecite{LuCH}]. However, experimentally
synthesized 9-7-9 and 7-9-7 AGNRHs lack translational symmetry, as
illustrated in Fig.~1(a). Because the wave-function decay length
of the TS is very short, 9-7-9 AGNRH segments do not exhibit
significant size effects on the TSs [\onlinecite{DJRizzo}], which
explains why predictions based on the Zak number remain valid for
9-7-9 and 7-9-7 AGNRH segments [\onlinecite{JiangJW}]. For wider
AGNRHs, such as 15-13-15, 21-19-21, and 27-25-27 segments, the
wave-function decay lengths of TSs increase with width, making Zak
number calculations insufficient. This motivates a real-space
analysis of the relationship between ESs of wider AGNR segments
possessing multiple end states (ESs) and IFs in wider AGNRHs,
which is critical for designing two-dimensional TS-based crystals
for novel quantum devices [\onlinecite{GuYW}].

A real-space approach involves solving the Schrodinger equation to
obtain eigenvalues and eigenfunctions of AGNRHs. Exact analytical
solutions exist only for triangular or rectangular GNR structures
within the one-band tight-binding model
[\onlinecite{Maalysheva}--\onlinecite{TalkachovA}]. Numerical
methods can alternatively determine the number of IFs in AGNR
heterojunctions [\onlinecite{LopezSancho}--\onlinecite{GuzmanM}].
However, distinguishing the locations and counts of IFs in
numerical calculations is challenging due to their zero-energy
modes. To overcome this, we propose applying a longitudinal
electric field to AGNRH segments, as illustrated in Fig.~1(b). The
resulting Stark effect not only resolves the number of IFs but
also clarifies their spatial locations.

This study has two main objectives. First, we aim to elucidate the
correlation between the ESs of isolated AGNR segments and IFs in
N-AGNR/$(N-2)$-AGNR/N-AGNR heterostructures using a real-space
formulation. While only 9-7-9 and 7-9-7 AGNRHs have been
experimentally realized via bottom-up synthesis
[\onlinecite{DJRizzo}], we extend our analysis to semiconducting
AGNRs in the N = 9, 15, 21, and 27 families. To explore the
evolution of terminal states of isolated AGNRs into IFs of AGNRHs,
we introduce a tunable inter-AGNR hopping parameter $t_{es}$ at
the junction sites (Fig.~1(b)), following the bulk-boundary
perturbation method [\onlinecite{GuzmanM}]. When $t_{es} = 0$, the
AGNRH decouples into three independent segments; varying $t_{es}$
effectively modifies the boundary conditions of each segment.
Sublattices are labeled A and B in Fig.~1(b) to emphasize the
chiral symmetry of the structure [\onlinecite{JiangJW}].

Second, we analyze charge transport through IFs in the Coulomb
blockade regime when zigzag-terminated ends of the AGNRH are
connected to electrodes (Fig.~1(c)). Tunneling current
measurements probe the spatial distribution of localized states,
providing essential information for designing one- and
two-dimensional 0D-TS crystals. We also tackle the challenging
calculation of nonlinear thermoelectric power in topological
double quantum dots (TDQDs) formed by AGNRH segments. By employing
an analytical solution for tunneling current in the Coulomb
blockade regime, we evaluate the influence of Coulomb interactions
on both thermal voltage and thermal current, providing a
comprehensive understanding of TDQD thermoelectric performance.

\begin{figure}[h]
\centering
\includegraphics[trim=1.cm 0cm 1.cm 0cm,clip,angle=0,scale=0.3]{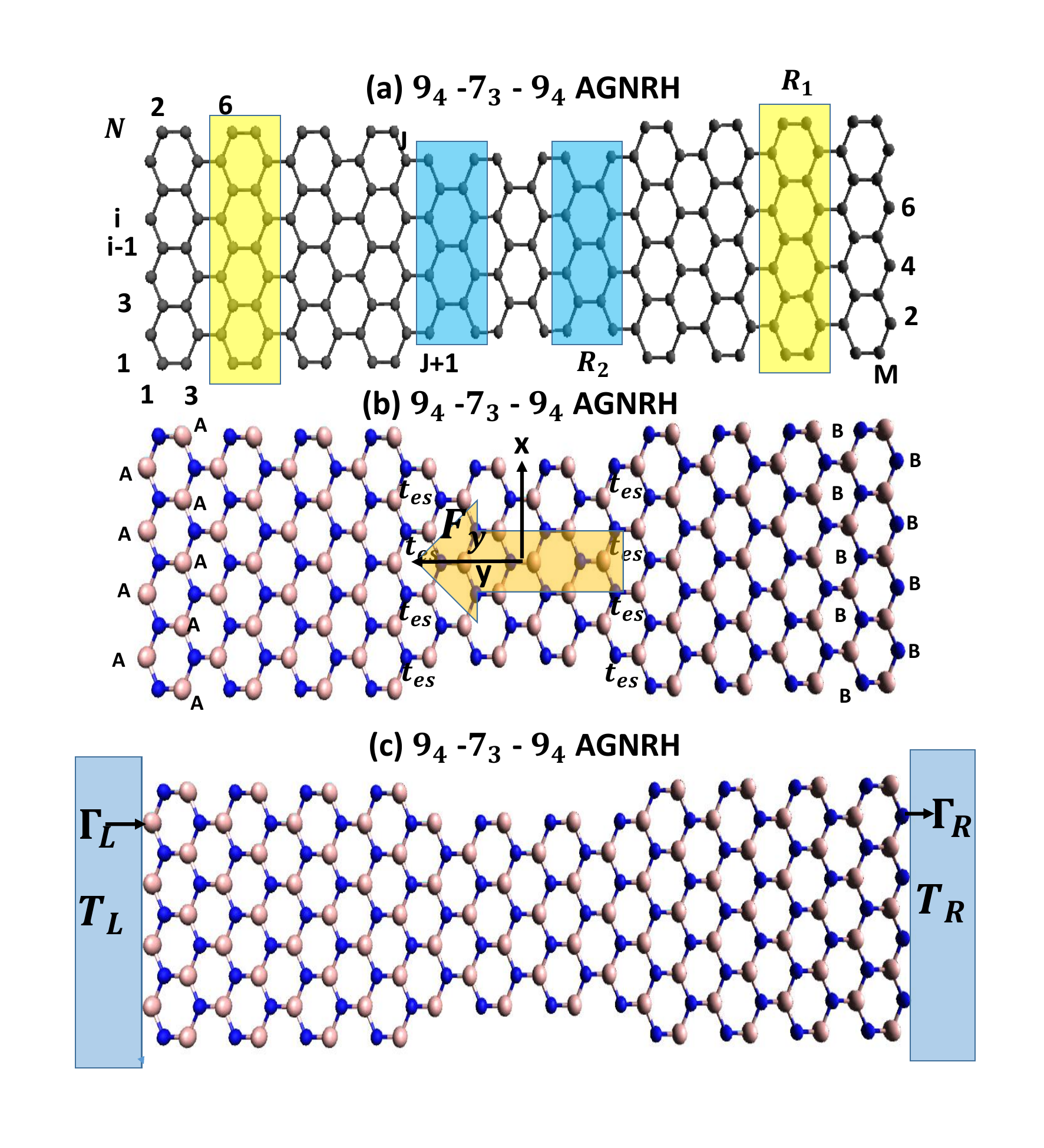}
\caption{ (a) Schematic illustration of a $9_w-7_x-9_y$ armchair
graphene nanoribbon heterostructure (AGNRH) composed of three AGNR
segments. The size of the AGNRH is characterized by ($N,M$), where
$N$ and $M$ denote the row and column numbers. $R_1$ and $R_2$
represent the unit cells (u.c.) of the 9-AGNR and 7-AGNR segments,
respectively. (b) A and B sublattice sites are indicated by white
and blue colors. The inter-AGNR electron hopping strengths
$t_{es}$ connect adjacent atoms between the 9-AGNR and 7-AGNR
segments. (c) AGNRH segment with zigzag edge terminations coupled
to left and right electrodes with equilibrium temperatures $T_L$
and $T_R$. $\Gamma_{L(R)}$  denote the tunneling rates for
electrons tunneling from the left (right) electrode into the
adjacent atoms at the zigzag edges.}
\end{figure}

\section{Calculation Methodology}
The system Hamiltonian, illustrated in Fig.~1(c), is written as:
$H=H_0+H_{GNR}$, where $H_0$ describes the Hamiltonian of the
electrodes and $H_{GNR}$ represents the AGNRH. The Hamiltonian of
$H_0$ is given by

\begin{small}
\begin{eqnarray}
H_0& = &\sum_{k,\sigma} \epsilon_k
a^{\dagger}_{k,\sigma}a_{k,\sigma}+
\sum_{k,\sigma} \epsilon_k b^{\dagger}_{k,\sigma}b_{k,\sigma}\\
\nonumber &+&\sum_{\ell,k,\sigma}
V^L_{k,\ell,j}d^{\dagger}_{\ell,j,\sigma}a_{k,\sigma}
+\sum_{\ell,k,\sigma}V^R_{k,\ell,j}d^{\dagger}_{\ell,j,\sigma}b_{k,\sigma}
+ h.c.,
\end{eqnarray}
\end{small}
where the first two terms describe the free electrons in the left
and right electrodes. The operators $a^{\dagger}_{k,\sigma}$
($b^{\dagger}_{k,\sigma}$) create an electron with momentum $k$
and spin $\sigma$ in the left (right) electrode, each with energy
$\epsilon_k$. The terms $V^L_{k,\ell,j=1}$ and $V^R_{k,\ell,j=M}$
describe the coupling between the electrode and its adjacent atoms
in the $\ell$-th row of the AGNRH. The electronic state of the GNR
is described using a tight-binding model with one $p_z$ orbital
per carbon atom [\onlinecite{NakadaK}-\onlinecite{WakabayashiK2}].
The AGNRH Hamiltonian $H_{GNR}$ is written as:
\begin{small}
\begin{eqnarray}
H_{GNR} &= &\sum_{\ell,j} E_{\ell,j,\sigma} d^{\dagger}_{\ell,j,\sigma}d_{\ell,j,\sigma}\\
\nonumber&-& \sum_{\ell,j}\sum_{\ell',j'} t_{(\ell,j),(\ell', j')}
d^{\dagger}_{\ell,j,\sigma} d_{\ell',j',\sigma} + h.c,
\end{eqnarray}
\end{small}
where $E_{\ell,j}$ denotes the on-site energy for the $p_z$
orbital at row ${\ell}$ and column $j$. Spin-orbit interaction is
neglected in this model. Graphene possesses exceptionally weak
spin orbit coupling and negligible hyperfine interaction, owing to
the dominant presence of $^{12}C$ atoms with zero nuclear spin
[\onlinecite{AllenMT}-\onlinecite{GerardotBD}].

The operators $d^{\dagger}_{\ell,j,\sigma}$ and
$d_{\ell,j,\sigma}$ create and annihilate electrons at the site
$\ell$,$j$, respectively. The hopping integral
$t_{(\ell,j),(\ell', j')}$ describes electron transfer between
sites $\ell$,$j$ and $\ell'$,$j'$. The tight-binding parameters
used for the AGNRH are $E_{\ell,j} = 0$ for all sites and
$t_{(\ell,j),(\ell',j')} = t_{pp\pi} = 2.7$~eV for nearest
-neighbor hopping. A perturbative hopping term $t_{es}$ is
introduced to probe the interaction between pairs of states with
opposite chirality, as illustrated in Fig.~1(b). The effect of a
longitudinal electric field $F_y$ is incorporated through an
additional potential $U_y = e F_y y$ added to $E_{\ell,j}$, where
$F_y = V_y/L_a$, with $V_y$ being the applied bias and $L_a$ the
length of the AGNRH.

To investigate the electron transport through the AGNRH junction,
the transmission coefficient ${\cal T}_{GNR}(\varepsilon)$ is
evaluated using ${\cal T}_{GNR}(\varepsilon) =
4Tr[\Gamma_{L}(\varepsilon)G^{r}(\varepsilon)\Gamma_{R}(\varepsilon)G^{a}(\varepsilon)]
$, where $\Gamma_{L(R)}(\varepsilon)$ denote the tunneling rates
of the left (right) electrode, and $G^{r(a)}(\varepsilon)$ are the
retarded (advanced) Green functions of the AGNRH
[\onlinecite{SunQF}-\onlinecite{Kuo5}]. In the tight-binding
formulation, $\Gamma_{\alpha}(\varepsilon)$ and Green's functions
are matrices. The expression for $\Gamma_{L(R)}(\varepsilon)$ is
derived from the imaginary part of the self-energies, denoted as
$\Sigma^r_{L(R)}(\varepsilon)$, and is given by
$\Gamma_{L(R)}(\varepsilon)=-\text{Im}(\Sigma^r_{L(R)}(\varepsilon))=\pi\sum_k|V^{L(R)}_{k,\ell,j=1(M)}|^2\delta(\varepsilon-\epsilon_k)$.
For simplicity, we adopt the wide-band limitation and assume
$\Gamma_{L(R)}(\varepsilon)$ to be energy-independent, denoted
simply as $\Gamma_{L(R)}$.

\section{Results and Discussion}

\subsection{End states of AGNR segments and interface states of AGNRH segments}

Because the end states (ESs) of an AGNR segment possess
zero-energy modes, it is difficult to resolve their degeneracies
through numerical calculations alone. Meanwhile, to experimentally
reveal the number of ESs, we apply a Stark effect induced by an
external electric field to lift the zero-energy modes and thereby
determine the number of ESs in AGNR segments with an $R_1$ unit
cell (u.c.), as shown in Fig.~1(a). Figure~2 displays the energy
levels of $N$-AGNR segments under a uniform electric field applied
along the armchair direction (defined as the $y$-direction). The
conduction- and valence subband edge states, $E_c$ and $E_v$,
exhibit red Stark shifts, while the localized ESs exhibit blue
Stark shifts. These localized states show a linear dependence on
the applied voltage $V_y$. The subband gap $\Delta_g = E_c - E_v$
decreases with increasing ribbon width for a family of AGNR, as
illustrated in Fig.~2.

Furthermore, the number of ESs differs among AGNR segments of
various widths. We find that 13-AGNR and 15-AGNR segments possess
two left ESs ($N_A = 2$) and two right ESs ($N_B = 2$). The
energies of the left ESs are labeled as $\Sigma_{c1}$ and
$\Sigma_{c2}$, while those of the right ESs are labeled as
$\Sigma_{v1}$ and $\Sigma_{v2}$. The subscripts $c$ and $v$ denote
states above and below the charge-neutral point (CNP),
respectively. These multiple ESs contrast with those in narrower
segments such as 7-AGNR and 9-AGNR, which each exhibit only one
left ES and one right ES ($N_A = N_B = 1$). For wider AGNRs, such
as 19-AGNR and 21-AGNR, we obtain $N_A = N_B = 3$, and for 25-AGNR
and 27-AGNR, we obtain $N_A = N_B = 4$. Based on the numerical
results in Fig.~2, the ES numbers follow the rule $N = 6N_{A(B)} +
1$ and $N = 6N_{A(B)} + 3$. The former corresponds to
semiconducting AGNRs with $N = 3p + 1$ and the latter to
semiconducting AGNRs with $N = 3p$, where $p$ is an integer.

\begin{figure}[h]
\centering
\includegraphics[trim=1.cm 0cm 1.cm 0cm,clip,angle=0,scale=0.3]{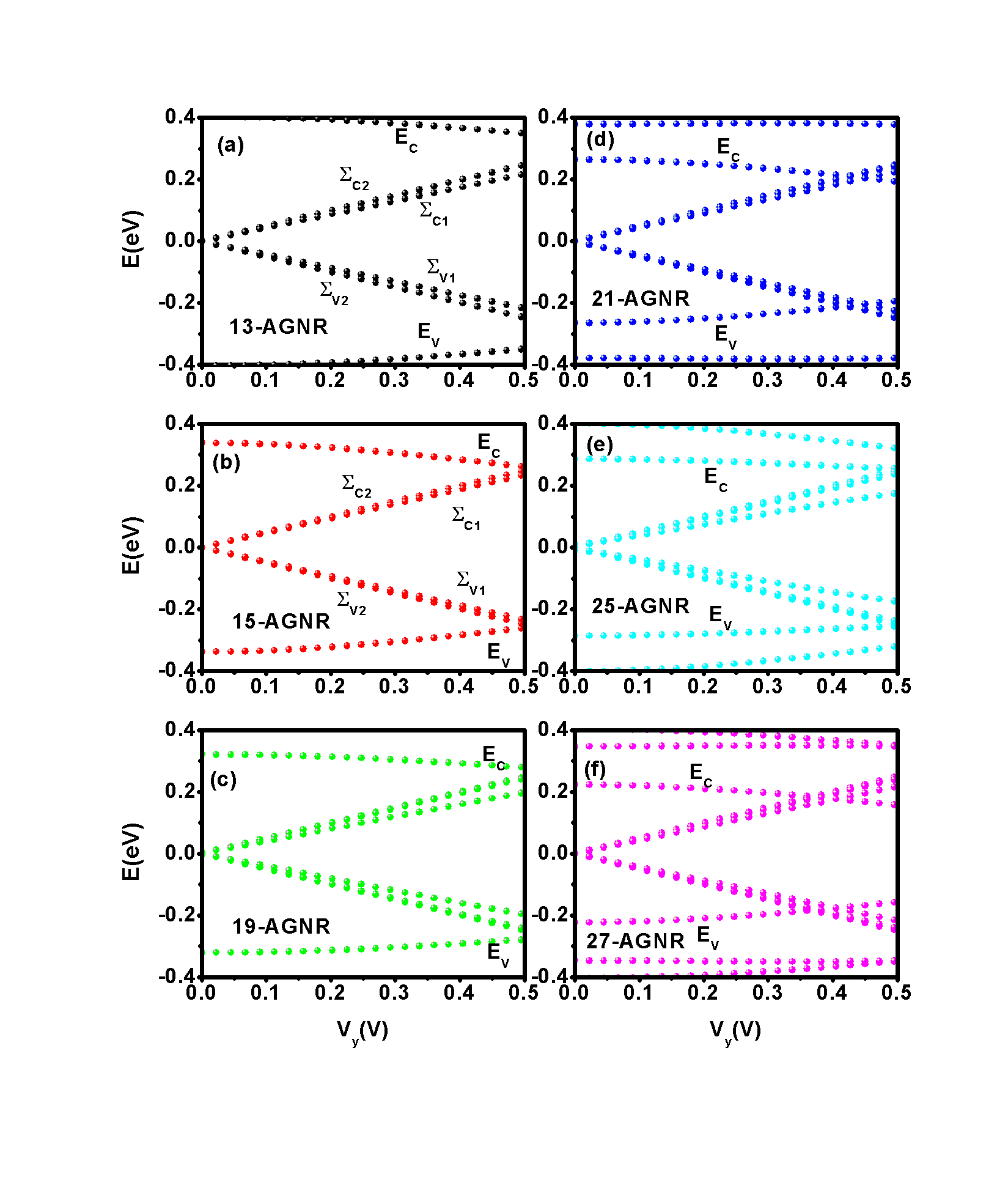}
\caption{Energy levels of armchair graphene nanoribbon (AGNR)
segments as functions of the applied voltage $V_y$. (a) 13-AGNR,
(b) 15-AGNR, (c) 19-AGNR, (d) 21-AGNR, (e) 25-AGNR, and (f)
27-AGNR segments. All segments have a length specified by $M =
96$, and their widths are characterized by $N$.}
\end{figure}

The Stark shift of ES energy levels in AGNR segments can also be
applied to AGNRH segments. For the N-AGNR/(N-2)-AGNR/N-AGNR
heterostructures considered in this work, we focus on the
semiconducting family with $N = 9, 15, 21$, and $27$. The outer
AGNR segments in these heterostructures belong to the $N = 3p$
family, while the central segment belongs to the $N = 3p + 1$
family. Figure~3 presents the energy spectra of these AGNRHs as
functions of the applied voltage $V_y$.

As shown in Fig.~3(a), the 9-7-9 AGNRH exhibits two low-energy
modes above the CNP and two below it. The modes labeled
$\Sigma_{c,1}$ and $\Sigma_{v,1}$ correspond to the left and right
ESs of the 9-7-9 AGNRH segment. The modes $\Sigma_{IF,c}$ and
$\Sigma_{IF,v}$ represent the left and right topological
symmetry-protected interface states (IFs) at the 9-7 and 7-9 AGNR
heterojunctions. At $V_y = 0.18$~V, the energy levels are
$\Sigma_{c,1} = 0.087$~eV, $\Sigma_{v,1} = -0.087$~eV,
$\Sigma_{IF,c} = 0.062$~eV, and $\Sigma_{IF,v} = -0.062$~eV.

For the wider AGNRH structures shown in Figs.~3(b)-3(d), only one
left IF and one right IF appear, even though the AGNR segments
contain multiple ESs. For example, in Fig.~3(b), the 15-13-15
AGNRH at $V_y = 0.18$~V exhibits six in-gap states: four ESs with
energies $\Sigma_{c,2} = 0.08932$~eV, $\Sigma_{v,2} =
-0.08932$~eV, $\Sigma_{c,1} = 0.08483$~eV, and $\Sigma_{v,1} =
-0.08483$~eV, and two IFs with energies $\Sigma_{IF,c} =
0.05972$~eV and $\Sigma_{IF,v} = -0.05972$~eV.

Similarly, the 21-19-21 AGNRH in Fig.~3(c) exhibits eight in-gap
states at $V_y = 0.18$~V: six ESs with energies $\Sigma_{c,3} =
0.08966$~eV, $\Sigma_{v,3} = -0.08966$~eV, $\Sigma_{c,2} =
0.08823$~eV, $\Sigma_{v,2} = -0.08823$~eV, $\Sigma_{c,1} =
0.0824$~eV, and $\Sigma_{v,1} = -0.0824$~eV, along with two IFs at
$\Sigma_{IF,c} = 0.05728$~eV and $\Sigma_{IF,v} = -0.05728$~eV.
The 27-25-27 AGNRH contains ten in-gap states (eight ESs and two
IFs). Compared with the in-gap states of the isolated N-AGNR
segments shown in Fig.~2, the N-AGNR/(N-2)-AGNR/N-AGNR
heterostructures exhibit two additional in-gap states arising from
the IFs. Thus, the Stark effect reveals not only the number of ESs
in AGNR segments but also the number of IFs in AGNRH structures.

\begin{figure}[h]
\centering
\includegraphics[trim=1.cm 0cm 1.cm 0cm,clip,angle=0,scale=0.3]{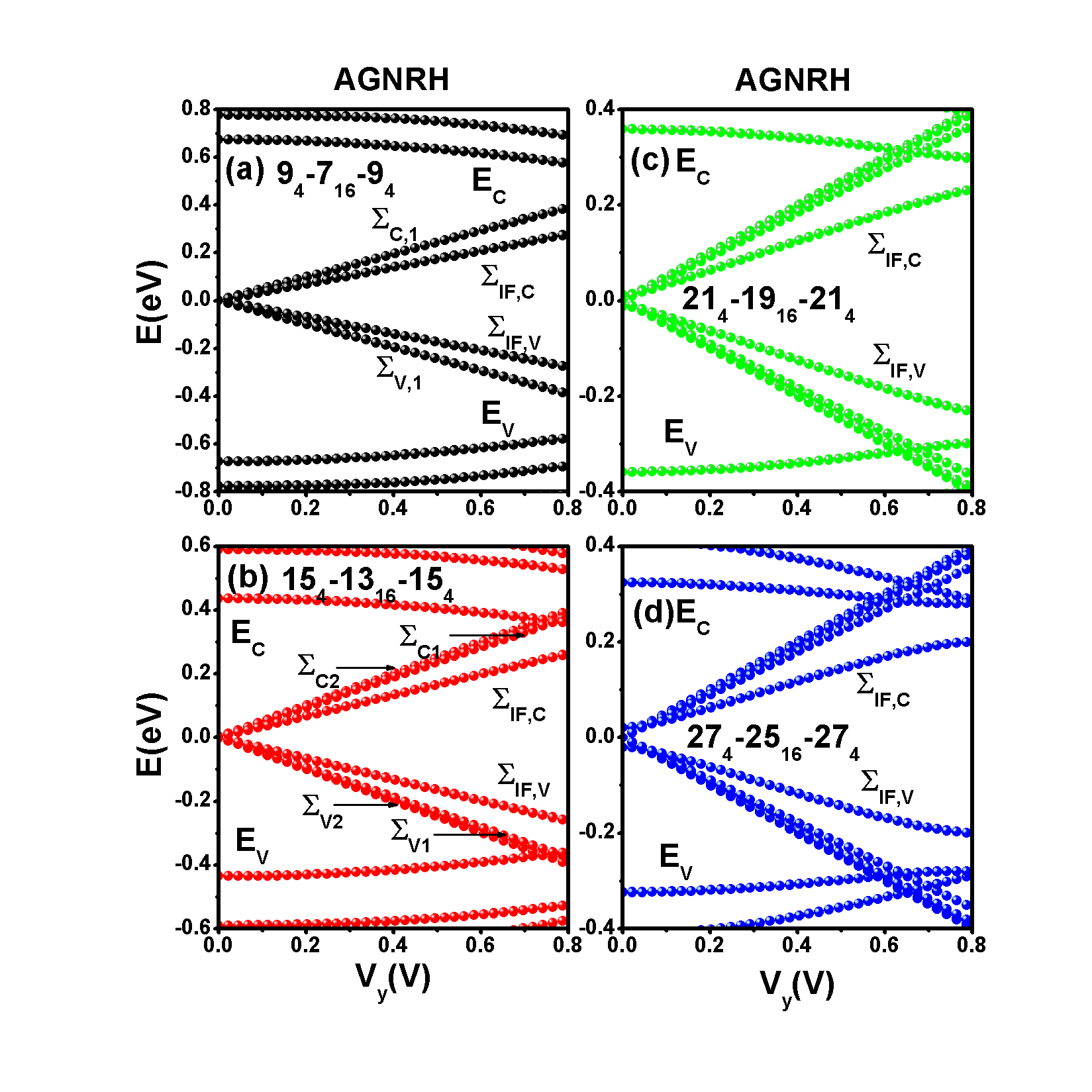}
\caption{Energy levels of armchair graphene nanoribbon
heterostructure (AGNRH) segments as functions of the applied
voltage $V_y$ for different widths.(a) $9-7-9$ AGNRH, (b)
$15-13-15$ AGNRH, (c) $21-19-21$ AGNRH, and (d) $27-25-27$ AGNRH.
Here, we adopt $w = y = 4$ for the outer AGNR segments with
$R_1$-type unit cells (u.c.) and $x = 16$ for the central AGNR
segment with an $R_2$-type unit cell (u.c.).}
\end{figure}

\subsection{Bulk boundary perturbation method}

Although the number of IFs in AGNR heterojunctions can be
determined using the winding number $Z$, its evaluation requires
momentum-space calculations~[\onlinecite{JiangJW}]. However, this
approach is not directly suitable for finite AGNRH segments. To
elucidate the correlation between the ESs of AGNR segments and the
IFs of AGNRH segments, we analyze the energy levels of the four
AGNRH structures shown in Fig.~3(a-d) as functions of the
inter-AGNR coupling parameter $t_{es}$. In Fig.~4(a), eight in-gap
energy levels appear at $t_{es} = 0$ because the 9-7-9 AGNRH is
decoupled into isolated left-, central-, and right-AGNR segments.
Each 9-AGNR segment with an $R_1$-type unit cell hosts $N_{L(R),A}
= 1$ and $N_{L(R),B} = 1$ ESs. The 7-AGNR segment, which has an
$R_2$-type unit cell, possesses $N_{C,A} = 2$ and $N_{C,B} = 2$
ESs. One of the $N_{C,A}$ ($N_{C,B}$) states is an additional
zero-energy mode with an extremely localized wave-function at
sublattice-A(B), denoted as $\Psi_{7,A}$ ($\Psi_{7,B}$). Because
the 9-AGNR and 7-AGNR segments have finite lengths of 4 and 6 unit
cells, respectively, these in-gap ESs hybridize and form bonding
and antibonding energy levels. However, the coupling between
$\Psi_{7,A}$ and $\Psi_{7,B}$ remains very weak.

As $t_{es}$ increases from 0 to the pristine hopping value $t =
2.7$~eV, four of the in-gap states exhibit linear $t_{es}$
dependence, forming the curves labeled $\Sigma_{AB,c}$ and
$\Sigma_{AB,v}$. These levels merge into the bulk bands at $t_{es}
= 0.45~t$. The origin of $\Sigma_{AB,c}$ and $\Sigma_{AB,v}$ is
the interaction between pairs of ESs with opposite chirality
through weak boundary perturbation (see also Fig.~5). The curves
labeled $\Sigma_{IF,c}$ and $\Sigma_{IF,v}$ correspond to the
bonding and antibonding combinations of the robust ESs
($\Phi_{7,A}$ and $\Phi_{7,B}$) in the 7-AGNR segment. These
states are weakly dependent on $t_{es}$. Specifically, we find
$\Sigma_{IF,c} = 0.04552$~eV ($\Sigma_{IF,v} = -0.04552$~eV) at
$t_{es} = 0$, and $\Sigma_{IF,c} = 0.02286$~eV ($\Sigma_{IF,v} =
-0.02286$~eV) at $t_{es} = t_{pp\pi} = 2.7$~eV. The evolution of
these levels with $t_{es}$ demonstrates that the robust ESs of the
7-AGNR segment give rise to the interface states of the 9-7-9
AGNRH. The terminal states of the 9-7-9 AGNRH, labeled
$\Sigma_{A,c}$ and $\Sigma_{B,v}$ near the CNP, are insensitive to
$t_{es}$.

For the 15-13-15 AGNRH shown in Fig.~4(b), we obtain 14 in-gap
levels at $t_{es} = 0$. The 15-AGNR segments possess $N_{L,A(B)} =
2$ and $N_{R,A(B)} = 2$, whereas the 13-AGNR segment has
$N_{C,A(B)} = 3$. As $t_{es}$ increases to $t = 2.7$~eV, eight ESs
move out of the bulk gap. The levels $\Sigma_{IF,c}$ and
$\Sigma_{IF,v}$ originate from the bonding and antibonding
combinations of the robust ESs of the 13-AGNR segment
($\Phi_{13,A}$ and $\Phi_{13,B}$). In this case, the robust ESs
exhibit a significantly longer decay length than those in the
9-7-9 AGNRH. In Figs.~4(c) and 4(d), 12 ESs and 16 ESs leave the
bulk gap in the 21-19-21 and 27-25-27 AGNRHs, respectively. The
21-19-21 AGNRH retains two IFs (forming $\Sigma_{IF,c}$ and
$\Sigma_{IF,v}$) and six ESs within the gap, while the 27-25-27
AGNRH contains two IFs and eight ESs. These results lead to the
general rule for $N$-AGNR/$(N-2)$-AGNR/$N$-AGNR heterostructures:
$N_{IF,L,A} = | N_{C,A} - N_{L,B} |$ and $N_{IF,R,B} = | N_{C,B} -
N_{R,A} |$. This relation shows that the number and type of IFs in
a symmetric
$N_{\text{out}}$-AGNR/$N_{\text{cen}}$-AGNR/$N_{\text{out}}$-AGNR
heterostructure with $\Delta N = (N_{\text{out}} -
N_{\text{cen}})/2 = 1$ are determined by the ES-number difference
and the chirality of the ESs in the AGNR segments with the maximal
ES number. As demonstrated in Appendix A, this rule is also valid
for $\Delta N \neq 1$, for example, $\Delta N = 3, 7, 9$.

\begin{figure}[h]
\centering
\includegraphics[trim=1.cm 0cm 1.cm 0cm,clip,angle=0,scale=0.3]{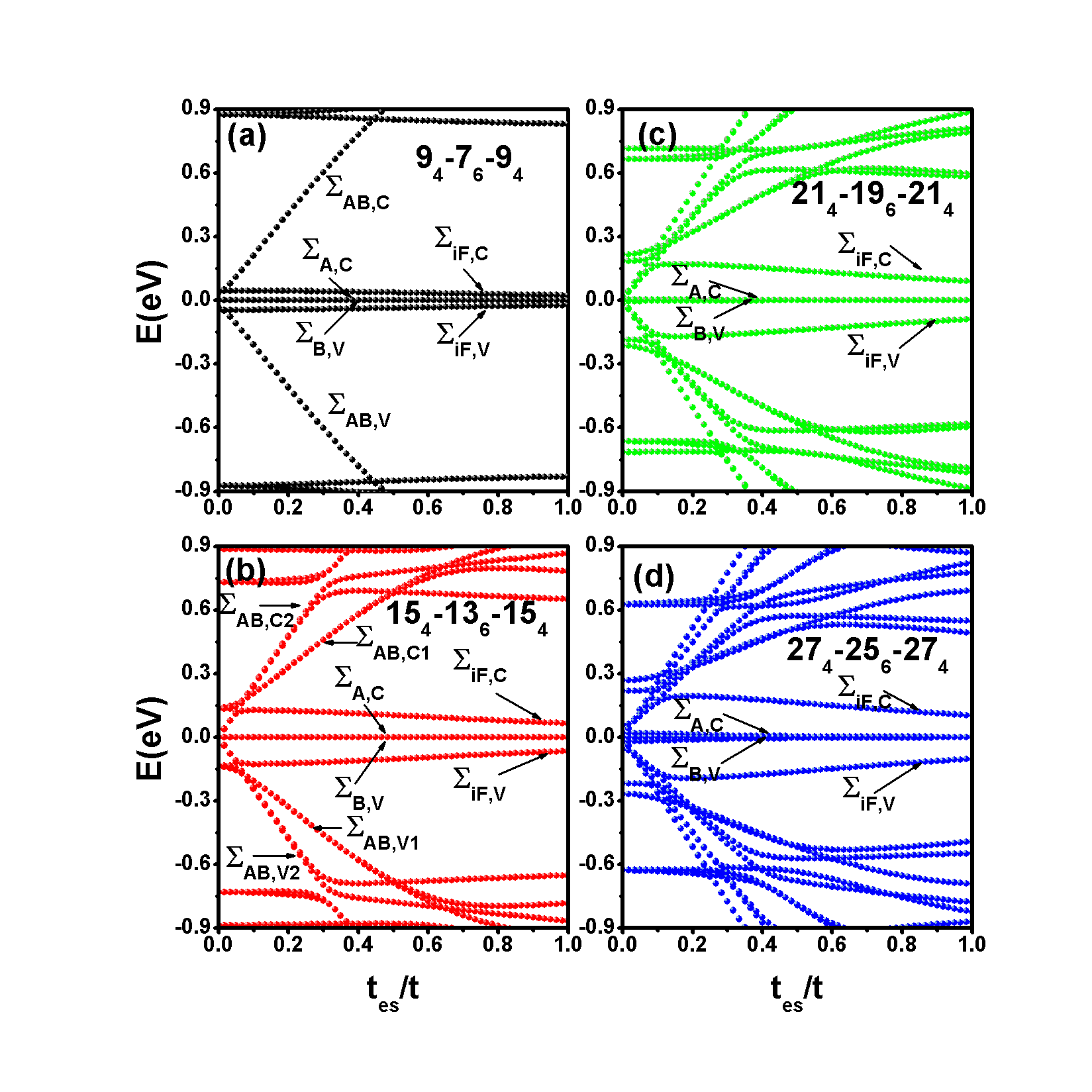}
\caption{Energy levels of four AGNRH segments with different
widths as functions of the inter-AGNR electron hopping strength
$t_{es}$. (a) $9_{4}-7_{6}-9_{4}$ AGNRH segment, (b)
$15_{4}-13_{6}-15_{4}$ AGNRH segment, (c) $21_{4}-19_{6}-21_{4}$
AGNRH segment and (d) $27_{4}-25_{6}-27_{4}$ AGNRH segment. The
parameter $t_{es}$is illustrated in Fig.~1(b).}
\end{figure}

Next, we consider the case of Fig.~4(b) as an example to
illustrate the charge (probability) density distribution of ESs
and IFs at different $t_{es}$ values, and to clarify why some ESs
shift out of the bulk gap, leaving only two IFs in the
N-AGNR/(N-2)-AGNR/N-AGNR structure. As shown in Figs.~5(a-c), the
AGNRH behaves as three weakly coupled segments at $t_{es} = 0.1t$.
In Fig.~5(a), the probability density of $\Sigma_{AB,c2}$ is
confined at the interface sites, reflecting interactions between
sublattice-A and sublattice-B states. The probability density of
$\Sigma_{AB,c1}$, shown in Fig.~5(b), also reflects
opposite-chirality interactions but is primarily distributed over
the outer 15-AGNR segments. By contrast, the probability density
of $\Sigma_{IF,c}$ is localized entirely within the central
13-AGNR segment (Fig.~5(c)).

At $t_{es} = 0.2t$, the distributions of $\Sigma_{AB,c2}$,
$\Sigma_{AB,c1}$, and $\Sigma_{IF,c}$ remain nearly unchanged. To
clarify the effect of $t_{es}$ on $\Sigma_{IF,c}$, we plot the
probability densities for $t_{es} = 0.5t$, $0.8t$, and $1.0t$ in
Figs.~5(g-i). As $t_{es}$ increases, the probability density of
$\Sigma_{IF,c}$ gradually extends into the outer 15-AGNR segments
while conserving normalization. Based on Figs.~4(b) and 5, we
conclude that the interface states of the N-AGNR/(N-2)-AGNR/N-AGNR
heterostructure originate from the ESs of the narrow $(N-2)$-AGNR
segment, which adopts the $R_2$-type unit cell. It is important to
note that the IF states exhibit a direction-dependent decay
length. Similar direction-dependent decay behavior has also been
reported for the two-dimensional topological interface states in
HgTe/CdTe heterostructures~[\onlinecite{ChangYC}].

\begin{figure}[h]
\centering
\includegraphics[angle=0,scale=0.25]{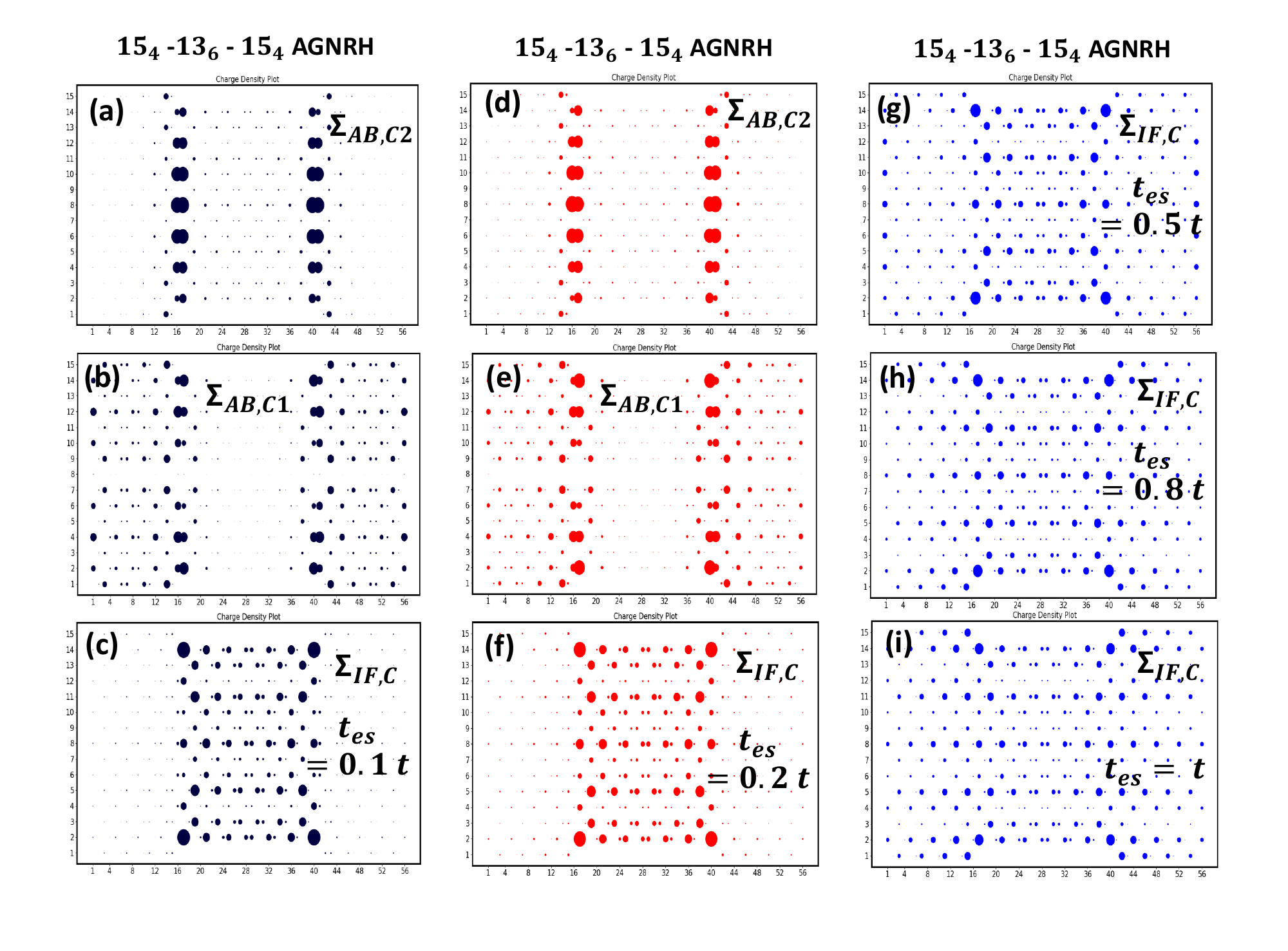}
\caption{(a-c) Probability densities corresponding to
$\Sigma_{AB,c2} = 0.24355$~eV, $\Sigma_{AB,c1} = 0.20466$~eV and
$\Sigma_{IF,c}= 0.12651 $~eV in the $15_{4}-13_6-9_{4}$ AGNRH
segment with $t_{es} = 0.1~t$. (d-f) Probability densities
corresponding to $\Sigma_{AB,c2} = 0.47859 $~eV, $\Sigma_{AB,c1} =
0.32919 $~eV and $\Sigma_{IF,c}= 0.12399 $~eV in the
$15_{4}-13_6-15_{4}$ AGNRH segment with $t_{es} = 0.2~t$. (g-i)
Probability densities of $\Sigma_{IF,c}$ in the
$15_{4}-13_6-9_{4}$ AGNRH segment for various values of
$t_{es}$:(g) $\Sigma_{IF,c}= 0.10343$~eV at $t_{es} = 0.5~t$, (h)
$\Sigma_{IF,c}= 0.07926$~eV at $t_{es} = 0.8~t$, and (i)
$\Sigma_{IF,c}= 0.06531 $~eV at $t_{es} = 1~t$.}
\end{figure}

\subsection{Interface states of AGNRH segments functioning as a single TDQD}

The bulk gap of the 9-7-9 AGNRH segment is larger than that of the
15-13-15, 21-19-21, and 27-25-27 AGNRH segments, which is
advantageous for suppressing thermal noise. However, the decay
length of its IF wave functions is too short to allow the
fabrication of gate electrodes. Increasing the width of AGNRH
segments leads to longer decay lengths of IF wave functions, as
reflected by the increased energy separation between
$\Sigma_{IF,c}$ and $\Sigma_{IF,v}$ shown in Fig.~4. Because the
bulk gaps of the 21-19-21 and 27-25-27 AGNRH segments are smaller
than that of the 15-13-15 AGNRH segment, we focus on the transport
properties of the interface states in the 15-13-15 AGNRH segment.
The calculated transmission coefficients for this structure are
shown in Fig.~6.

Figure~6(a) presents the transmission coefficient ${\cal
T}_{GNR}(\varepsilon)$ of a $15_4$-$13_6$-$15_4$ AGNRH with zigzag
edge terminations for different values of $t_{es}$. As $t_{es}$
decreases, the ${\cal T}_{GNR}(\varepsilon)$ spectrum splits into
two resonant peaks, each reaching a maximum value of one. These
two peaks clearly demonstrate that the 15-13-15 AGNRH segment
hosts two nondegenerate resonant channels. The peak positions
shift away from the CNP as $t_{es}$ decreases. This behavior
arises because reducing $t_{es}$ effectively increases the barrier
height experienced by the IFs, strengthening their confinement.
Consequently, the peak widths also become narrower. In the context
of $t_{es} \neq t$, one may consider employing STM techniques to
manipulate the electron-hopping strengths at the interface
sites~[\onlinecite{DJRizzo}].

Figure~6(b) shows ${\cal T}_{GNR}(\varepsilon)$ for different
staggered potentials $\delta_A = -\delta_B = \delta$ induced by
the two-dimensional substrates supporting the outer 15-AGNR
segments. Increasing $\delta$ enhances quantum confinement, and
once the 15-AGNR segments behave as potential barriers, strong
backward electron scattering is expected. However, since strain
engineering can be used to minimize the staggered potential
[\onlinecite{GiovannettiG}-\onlinecite{ZhaoJ}], we neglect
substrate-induced staggered potentials in the following
discussion.

In Fig.~6(c), we show ${\cal T}_{GNR}(\varepsilon)$ for the
$15_4$-$13_6$-$15_4$ AGNRH segment for various coupling strengths
$\Gamma$, corresponding to different electrode materials. While
the resonant peak positions remain unchanged, the peak widths vary
according to the coupling between the electrodes and the edge
atoms (i.e., variation of $\Gamma_L=\Gamma_R=\Gamma$). These
results further highlight the topologically protected and
symmetric localization characteristics of the IFs.

In Figs.~6(a)-(c), the length of the $15_4-13_6-15_4$ segment is
fixed. To examine size effects, we also calculate ${\cal
T}_{GNR}(\varepsilon)$ for $15_w$-$13_6$-$15_y$ AGNRHs with
symmetrically varied outer segment lengths ($w = y$). Because the
IFs are strongly localized, increasing $w$ reduces the effective
coupling strength between the electrodes and the IFs. As a result,
the resonant peaks become narrower with increasing 15-AGNR segment
length. Overall, the results in Fig.~6 demonstrate that the IFs in
a 15-13-15 AGNRH function as a topological double quantum dot
(TDQD). Each quantum dot hosts a single energy level that is well
isolated from bulk states by an effective bulk gap on the order of
one electron-volt (see Fig.~3(b)).

\begin{figure}[h]
\centering
\includegraphics[trim=1.cm 0cm 1.cm 0cm,clip,angle=0,scale=0.3]{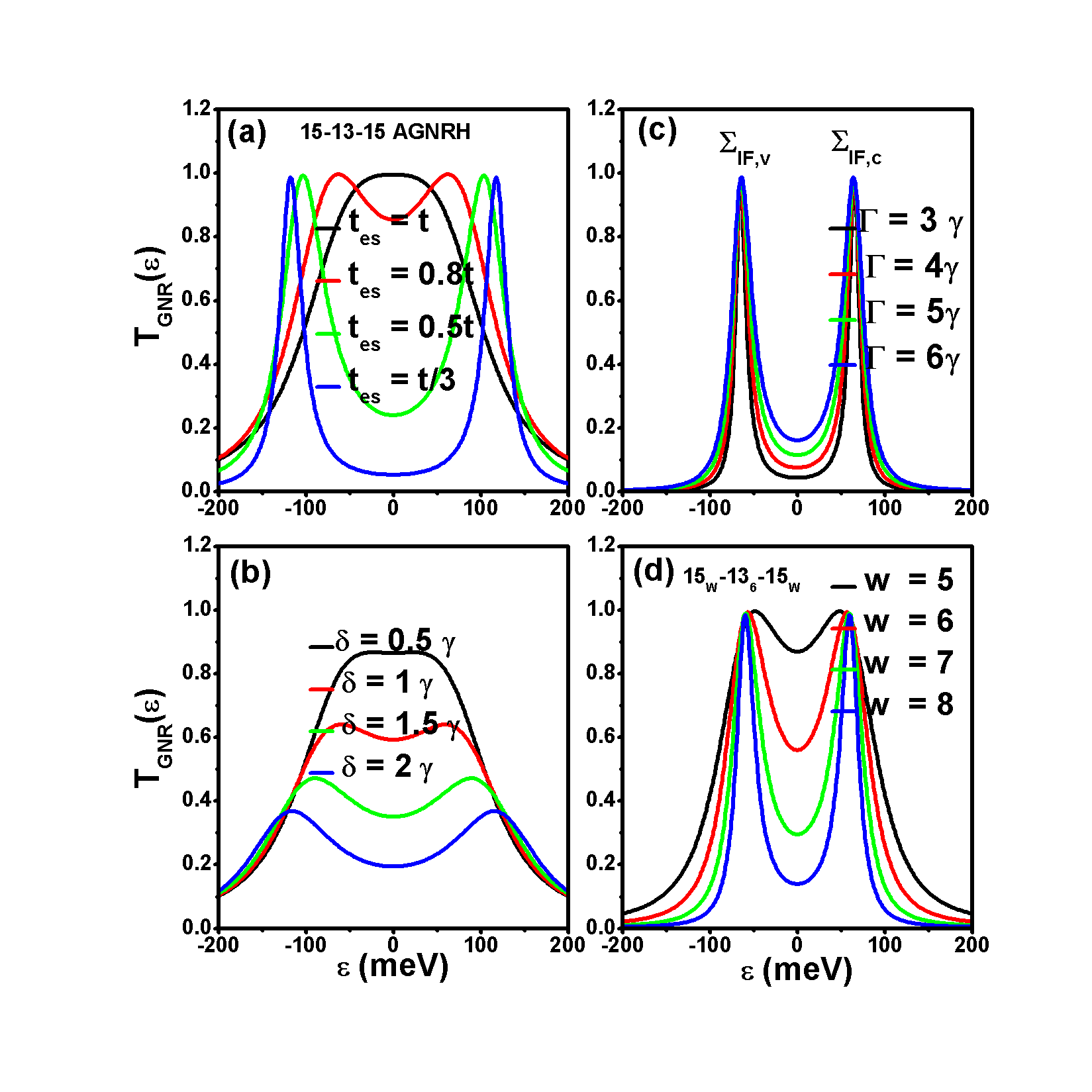}
\caption{Transmission coefficient ${\cal T}_{GNR}(\varepsilon)$ of
the $15_w-13_x-15_y$ AGNRH segment with zigzag-terminated edge
structures coupled to electrodes.(a) ${\cal T}_{GNR}(\varepsilon)$
of the $15_4-13_6-15_4$ AGNRH segment for different values of
$t_{es}$ at $\Gamma_L = \Gamma_R = \Gamma = 2.7$~eV. (b) ${\cal
T}_{GNR}(\varepsilon)$ of the $15_4-13_6-15_4$ AGNRH segment with
a staggered sublattice potential $\delta$, applied to the outer
AGNR segments, for $\Gamma=2.7$~eV and $t_{es} = t$.  (c) ${\cal
T}_{GNR}(\varepsilon)$ of the $15_4-13_6-15_4$ AGNRH segment with
$t_{es} = t$ and $\delta = 0$ for different values of $\Gamma$,
corresponding to various electrode materials. (d) ${\cal
T}_{GNR}(\varepsilon)$ of the $15_w-13_6-15_w$ AGNRH segment for
different values of $w$, with $t_{es} = t$, $\Gamma = 2.7$~eV and
$\delta = 0$. A parameter of $\gamma = 90$~meV is used in panels
(b) and (c).}
\end{figure}

\subsection{Nonlinear thermoelectric power of TDQD}

The transmission coefficients ${\cal T}_{GNR}(\varepsilon)$ shown
in Fig.~6 were calculated within a single-particle framework.
However, evaluating ${\cal T}_{GNR}(\varepsilon)$ in the Coulomb
blockade regime is extremely challenging
[\onlinecite{SolsF}--\onlinecite{ZhangJain}]. When intra-site
Coulomb interactions are included in the Hamiltonian of Eq.~(2),
current theoretical approaches remain limited to mean-field
treatments, i.e., one-particle approximations
[\onlinecite{JacobsePH}]. Since our interest lies in the
low-energy modes near the CNP, we employ an extended Anderson
model that incorporates effective intra-dot and inter-dot Coulomb
interactions to investigate tunneling through the TDQD. The
effective Hamiltonian is given by $H_{\text{eff}} = H_{\text{SD}}
+ H_{\text{TDQD}}$, where $H_{\text{SD}}$ describes the source and
drain electrodes and $H_{\text{TDQD}}$ represents the TDQD:

\begin{small}
\begin{eqnarray}
& &H_{TDQD}\\ \nonumber &= &\sum_{j=L,R,\sigma}E_{j}
c^{\dagger}_{j,\sigma}c_{j,\sigma}-t_{x} (c^{\dagger}_{R,\sigma}
c_{L,\sigma} + c^{\dagger}_{L,\sigma} c_{R,\sigma})\\ \nonumber
&+&\sum_{j=L,R}U_j~n_{j,\sigma}n_{j,-\sigma}
+\frac{1}{2}\sum_{j\neq\ell,\sigma,\sigma'}U_{j,\ell}~n_{j,\sigma}n_{\ell,\sigma'},
\end{eqnarray}
\end{small}

Here, $E_j$ is the spin-independent energy level of dot $j$, and
$U_j = U_{L(R)} = U_0$ and $U_{j,\ell}=U_{LR}=U_1$ denote the
intra-dot and inter-dot Coulomb interactions, respectively. The
operator $n_{j,\sigma}=c^\dagger_{j,\sigma}c_{j,\sigma}$ is the
number operator, and $t_x$ is the interdot hopping amplitude. For
15-13-15 AGNRH segments, both end states and interface states
emerge within the middle gap and are well separated from the
conduction and valence subbands by a band gap of approximately 1
eV [Fig. 3(b)]. This large separation suppresses optical-phonon
assisted transport due to the phonon bottleneck effect
[\onlinecite{FangJW}]; therefore, thermal noise effects arising
from bound-to-continuum states can be safely neglected in Eq. (3).
Noting that the phonon mean free path in graphene nanostructures
can be reduced to 10~nm
[\onlinecite{ZuevYM}-\onlinecite{HsiehYP}]. Since the lengths of
the AGNRH segments considered in this work are smaller than this
scale, electron-acoustic phonon scattering can be safely ignored.

The Green-function technique provides a powerful framework for
calculating tunneling currents in strongly correlated
nanostructures under nonequilibrium conditions
[\onlinecite{DavidK1}--\onlinecite{Kuo6}]. Using the
equation-of-motion method, the tunneling current from the left
(right) electrode through the TDQD with electron-electron
interactions is given by

\begin{eqnarray}
& &J_{L(R)}(V_{a},\Delta T)\\ \nonumber &=&\frac{2e}{h}\int
{d\varepsilon}~ {\cal
T}_{LR(RL)}(\varepsilon)[f_L(\varepsilon)-f_R(\varepsilon)],
\end{eqnarray}

The Fermi-Dirac distribution of electrode $\alpha$ is defined as
$f_{\alpha}(\varepsilon) =
1/({\exp[(\varepsilon-\mu_{\alpha})/k_BT_{\alpha}]+1}$),with
chemical potentials $\mu_{L(R)}=\mu \pm eV_a/2$ under an applied
bias of $+V_a/2$ and $-V_a/2$. For convenience, we set the Fermi
energy $\mu=0$. The temperature bias is defined as $\Delta
T=T_L-T_R$. A closed-form expression for ${\cal
T}_{LR}(\varepsilon)$ is given in Ref.~[\onlinecite{Kuo6}]:

\begin{small}
\begin{eqnarray}
& &{\cal T}_{LR}(\epsilon)/(4t^2_{x}\Gamma_{e,L}\Gamma_{e,R})=\frac{C_{1} }{|\epsilon_L\epsilon_R-t^2_{x}|^2} \nonumber \\
&+& \frac{C_{2} }{|(\epsilon_L-U_{LR})(\epsilon_R-U_R)-t^2_{x}|^2} \nonumber \\
&+& \frac{C_{3} }{|(\epsilon_L-U_{LR})(\epsilon_R-U_{LR})-t^2_{x}|^2} \label{TF} \\
\nonumber &+&
\frac{C_{4} }{|(\epsilon_L-2U_{LR})(\epsilon_R-U_{LR}-U_R)-t^2_{x}|^2}\\
\nonumber &+& \frac{C_{5} }{|(\epsilon_L-U_{L})(\epsilon_R-U_{LR})-t^2_{x}|^2}\\
\nonumber &+& \frac{C_{6}
}{|(\epsilon_L-U_L-U_{LR})(\epsilon_R-U_R-U_{LR})-t^2_{x}|^2}\\
\nonumber &+&
\frac{C_{7} }{|(\epsilon_L-U_L-U_{LR})(\epsilon_R-2U_{LR})-t^2_{x}|^2}\\
\nonumber
 &+&
\frac{C_{8} }{|(\epsilon_L-U_L-2U_{LR})(\epsilon_R-U_R-2U_{LR})-t^2_{x}|^2}, \\
\nonumber
\end{eqnarray}
\end{small}
Here, $\epsilon_L = \varepsilon - E_L + i\Gamma_{e,L}$ and
$\epsilon_R = \varepsilon -E_R + i\Gamma_{e,R}$; $\Gamma_{e,L(R)}$
is the tunneling rate determined by the dot-electrode coupling.
The eight terms above correspond to the eight possible TDQD
occupation configurations encountered by an incoming spin-$\sigma$
electron. The configuration probabilities $C_m$ are expressed as

\begin{small}
\begin{eqnarray}
C_{1}&=&1-N_{L,\sigma}-N_{R,\sigma}-N_{R,-\sigma}+ \langle
n_{R,\sigma}n_{L,\sigma}\rangle \nonumber \\ &+&\langle
n_{R,-\sigma}n_{L,\sigma}\rangle+\langle
n_{R,-\sigma}n_{R,\sigma}\rangle-\langle
n_{R,-\sigma}n_{R,\sigma} n_{L,\sigma} \rangle \nonumber \\
C_{2}&=&N_{R,\sigma}-\langle n_{R,\sigma} n_{L,\sigma}\rangle
-\langle n_{R,-\sigma} n_{R,\sigma}\rangle \nonumber \\ &+&\langle
n_{R,-\sigma} n_{R,\sigma} n_{L,\sigma}\rangle \nonumber \\
C_{3}&=&N_{R,-\sigma}-\langle n_{R,-\sigma} n_{L,\sigma}\rangle
-\langle n_{R,-\sigma}n_{R,\sigma}\rangle \nonumber \\ &+&\langle
n_{R,-\sigma}n_{R,\sigma} n_{L,\sigma} \rangle \nonumber \\
C_{4}&=&\langle n_{R,-\sigma}n_{R,\sigma}\rangle-\langle
n_{R,-\sigma}n_{R,\sigma} n_{L,\sigma}\rangle \nonumber\\
C_{5}&=&N_{L,\sigma}- \langle n_{R,\sigma}n_{L,\sigma}\rangle
-\langle n_{R,-\sigma} n_{L,\sigma}\rangle \nonumber \\ &+&\langle
n_{R,-\sigma}n_{R,\sigma} n_{L,\sigma}\rangle \nonumber \\
C_{6}&=&\langle n_{R,\sigma} n_{L,\sigma}\rangle -\langle
n_{R,-\sigma}n_{R,\sigma} n_{L,\sigma}\rangle \nonumber \\
C_{7}&=&\langle n_{R,-\sigma}n_{L,\sigma}\rangle -\langle
n_{R,-\sigma}n_{R,\sigma} n_{L,\sigma}\rangle \nonumber \\
C_{8}&=&\langle n_{R,-\sigma}n_{R,\sigma}n_{L,\sigma} \rangle
\nonumber,
\end{eqnarray}
\end{small}
Here, $N_{\ell,\sigma}$ is the single-particle occupation at site
$\ell$. Two-particle and three-particle correlation functions are
also included, and the full set is solved self-consistently so
that $\sum_m C_m = 1$, ensuring probability conservation. The
reverse-bias current is obtained by exchanging the indices of
${\cal T}_{LR}(\varepsilon)$. Note that the transmission formula
is valid only for temperatures above the Kondo temperature
[\onlinecite{MadhavanV}--\onlinecite{GoldhaberG}].

Although many-body effects in voltage-driven transport have been
widely explored, their impact on electrical power generation in
nanostructures driven by a temperature bias--both in
quasi-equilibrium and far-from-equilibrium regimes--remains
insufficiently understood
[\onlinecite{ChenG}--\onlinecite{SobrinoN}]. Here, we investigate
these many-body effects for thermoelectric generators formed by
the topological states (TSs) of AGNRHs. In realistic experimental
conditions, the left (right) gate voltage not only tunes the
energy level of the left (right) TQD but also affects the right
(left) TQD; therefore, the left and right TQD energy levels are
modulated as $E_L = -(\eta_1
eV_{L,g}+(1-\eta_1)eV_{R,g})+\eta_L~eV_{a}$ and $E_R =
-((1-\eta_2)eV_{L,g}+\eta_2V_{R,g})+\eta_R~eV_{a}$, where $\eta_1$
and $\eta_2$ represent the effects of intra- and interdot Coulomb
interactions [\onlinecite{DasSarma}] and $\eta_{L(R)}$
($|\eta_{L(R)}|<1/2$) describe the orbital shifts induced by $V_a$
(see Fig. (3)). We take $U_0=90$~meV and $U_1=25$~meV, consistent
with their weak dependence on AGNR lengths [\onlinecite{Kuo5}]. In
this study, we use $\eta_1 = \eta_2 = 0.8$ and a small orbital
offset $\eta_L = - \eta_R = 0.1$. The tunneling rates
$\Gamma_{e,L}$ and $\Gamma_{e,R}$ and the interdot hopping $t_x$
are treated as tunable parameters controlled by the 13- and
15-AGNR segment lengths.

To analyze nonlinear temperature-driven power generation, we
define the optimal electrical power as $\Omega_{op} =
-J_{op}(V_{op},\Delta T)\times V_{op}(\Delta T)$, where
$J_{op}(V_{op},\Delta T)$ and $V_{op}(\Delta T)$ denote the
optimized thermal current and thermal voltage, respectively
[\onlinecite{JiangJH},\onlinecite{FastJ}]. The thermal voltage
$V_{op}(\Delta T)$ is obtained by maximizing the electrical power
$\Omega = -J_{L(R)}(V_{a},\Delta T)\times V_{a}$ with respect to
$V_{a}$ [\onlinecite{FastJ}]. This optimization is highly
nontrivial in the Coulomb blockade regime under nonequilibrium
conditions; thus, many theoretical studies avoid correlated
transport and power optimization altogether
[\onlinecite{AzemaJ}--\onlinecite{SanchezR}]. The electrode
temperatures are expressed as $T_L = T_0 \pm \Delta T/2$ and $T_R
= T_0 \mp \Delta T/2$, where $T_0=(T_L+T_R)/2$ is the average
temperature.

\textbf{1. Quasi-equilibrium and Out-of-equilibrium scenarios}\\
Figures 7(a)-7(c) show the charge stability diagram ($N_t$),
electrical conductance ($G_e$), and Seebeck coefficient ($S$) as
functions of the two gate voltages $V_{L,g}$ and $V_{R,g}$ at
$T_0=12$ K. In Fig. 7(a), the charge stability diagram gives the
total charge number $N_t = N_L + N_R =
\sum_{\sigma}(N_{L,\sigma}+N_{R,\sigma})$. Nine distinct charge
configurations ($N_L,N_R$) are observed, corresponding to TDQD
occupancies of zero to four electrons. These results are
consistent with experimental charge stability diagrams reported
for other serial DQD systems at low temperature
[\onlinecite{DasSarma}--\onlinecite{EichM}]. In Fig. 7(b), the
electrical conductance exhibits eight peaks aligned with the
charge-transition lines in Fig. 7(a). These peaks indicate that
the TDQD provides eight resonant transport channels under weak
coupling ($\Gamma_{e,t} = t_x = 1$~meV). The first peak ($P_1$)
corresponds to the lowest-energy TDQD state, while the last peak
($P_8$) corresponds to the highest-energy
state[\onlinecite{Kuo6}]. The Seebeck coefficient in Fig. 7(c)
exhibits a bipolar behavior, with negative (positive) values
associated with electron (hole-like) transport. The $S$ spectra
thus encode both the thermal transport characteristics and the
charge stability structure of the TDQD. The quantities $G_e$ and
$S$ are expressed in units of $G_0 = 2e^2/h = 77.5~\mu$S and
$k_B/e = 86.25~\mu$V/K, respectively. These results represent the
quasi-equilibrium regime of TDQD thermoelectric transport.

\begin{figure}[h]
\centering
\includegraphics[trim=1.cm 0cm 1.cm 0cm,clip,angle=0,scale=0.3]{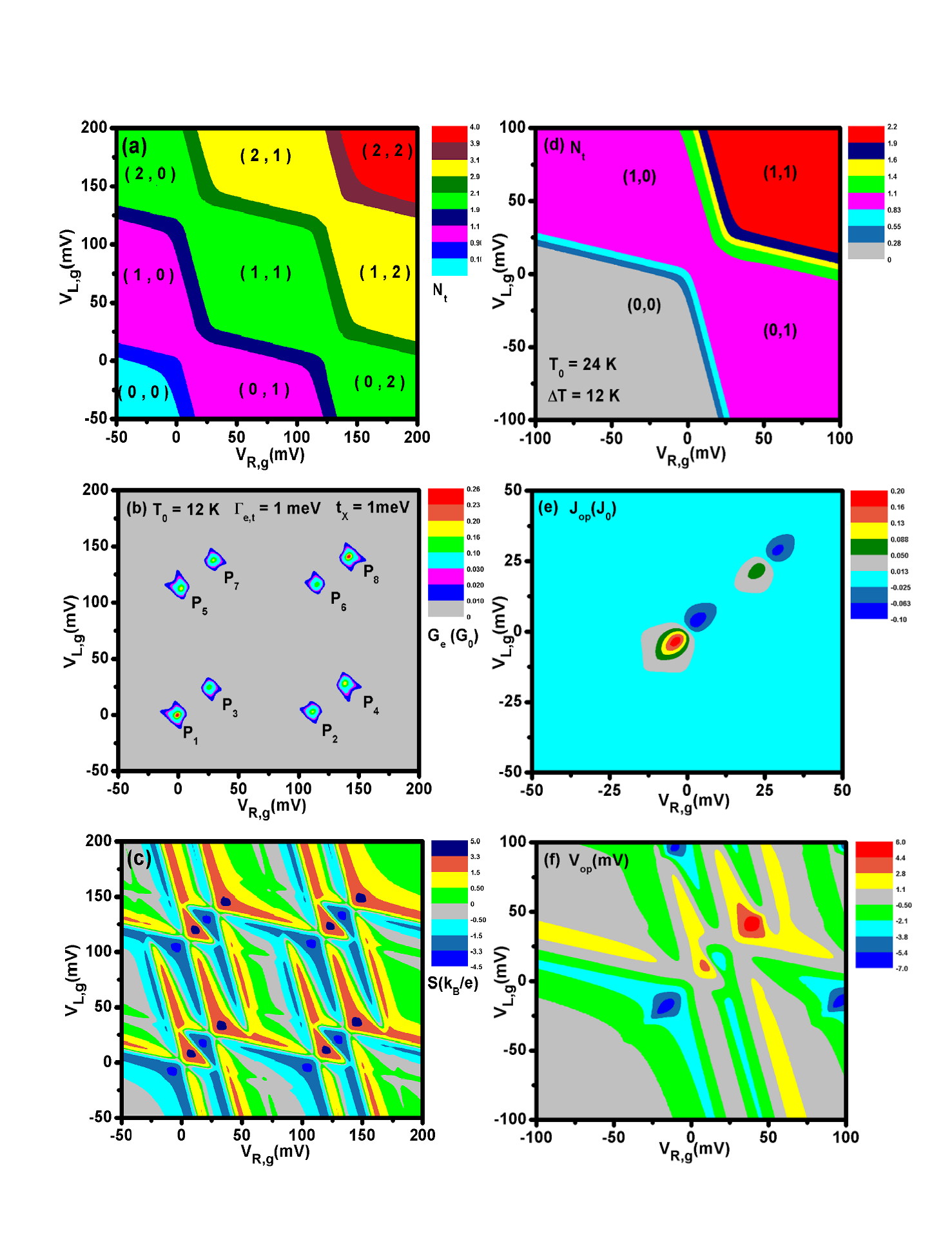}
\caption{(a) Charge stability diagram ($N_t$), (b) electrical
conductance ($G_e$), and (c) Seebeck coefficient ($S$) as
functions of gate voltages $V_{L,g}$ and $V_{R,g}$ at $T_0=12$ K
with $\Delta T=0$. (d) Total occupation number $N_t$, (e)
optimized thermal current $J_{op}(\Delta T)$, and (f) optimized
thermal voltage $V_{op}(\Delta T)$ at $T_0=24$ K with $\Delta
T=12$ K. A weak interdot coupling of $t_x=\Gamma_{e,t}=1$ meV is
used.}
\end{figure}

To examine the out-of-equilibrium regime, Figs. 7(d)-7(f) present
the total charge number $N_t$, optimized thermal current $J_{op}$,
and optimized thermal voltage $V_{op}$ at average temperature
$T_0=24$ K with a thermal bias $\Delta T=12$ K. In Fig. 7(d), the
charge-transition region between (0,1) and (1,1) is broader than
that between (1,0) and (1,1), reflecting the asymmetric electrode
temperatures ($T_L=30$ K, $T_R=18$ K). Unlike $G_e$, the optimized
thermal current $J_{op}$ [Fig. 7(e)] displays a bipolar dependence
on the symmetric gate voltage $V_{L,g}=V_{R,g}=V_g$, with
vanishingly small values at the conductance peaks. This indicates
that maximum $J_{op}$ does not occur when TDQD energy levels align
with the Fermi energy. The optimized thermal voltage $V_{op}$
[Fig. 7(f)] also shows bipolar behavior but with broader features
than the Seebeck coefficient. Although $J_{op}$ and $V_{op}$
exhibit opposite Coulomb oscillations with respect to $V_g$, the
optimized output power $\Omega_{op}$ remains positive.

\textbf{2. Comparison between noninteracting and interacting cases
}\\ Many previous studies predicted the power output of nanoscale
thermoelectric generators without considering electron-electron
Coulomb interactions
[\onlinecite{Hershfield}--\onlinecite{RaymondJH}]. To clarify the
role of Coulomb interactions and temperature bias on $V_{op}$ and
$J_{op}$ at room temperature, Fig. 8 shows $V_{op}$, $J_{op}$, and
$\Omega_{op}$ as functions of gate voltage $V_g$ and temperature
bias $\Delta T$ at $T_0 = 288$ K. $V_g$ is varied from $-100$ mV
to 0, and $\Delta T$ from 0 to 240 K. Fig. 8(a)-8(c) correspond to
the noninteracting case, and Fig. 8(d)-8(f) to the interacting
case. As seen in Fig. 8(a) and Fig. 8(d), $V_{op}$ differs only
slightly, mainly near $V_g \geq -10$~mV: without interactions,
$V_{op}(V_g = 0) = 0$ for all $\Delta T$ due to electron-hole
symmetry, whereas interactions lift this symmetry, giving small
but finite $V_{op}$ at $V_g = 0$. In contrast, $J_{op}$ shows
large differences: the noninteracting maximum $J_{op,\max} =
0.566$ at $V_g = -25$~mV and $\Delta T = 240$~K [Fig.8 (b)] is
significantly overestimated compared with the interacting value
$J_{op,\max} = 0.418$ at $V_g = -32$~mV [Fig. 8(e)]. In the (0,0)
configuration, the Coulomb blockade suppresses $J_{op,\max}$ but
has little effect on $V_{op,\max}$. The behavior of $V_{op}$
resembles that of the Seebeck coefficient $S$, which is largely
insensitive to electron Coulomb interactions [\onlinecite{Kuo9}].

\begin{figure}[h]
\centering
\includegraphics[angle=0,scale=0.3]{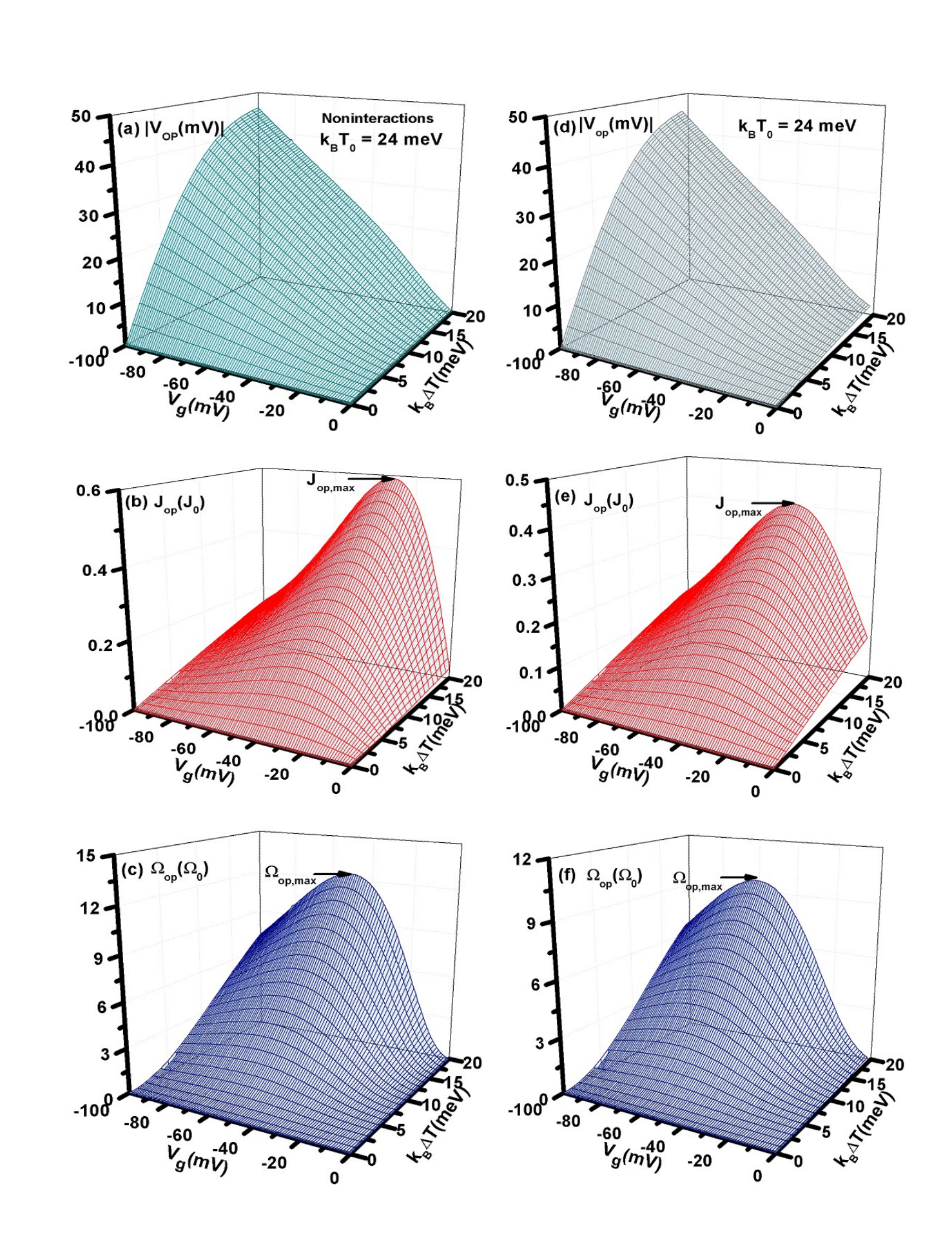}
\caption{(a-c) Noninteracting case: optimized thermal voltage
$V_{op}$, thermal current $J_{op}$, and electrical power
$\Omega_{op}$ as functions of symmetric gate voltage $V_{L,g} =
V_{R,g} = V_g$ and temperature bias $\Delta T$ at $T_0 = 288$ K.
(d-f) Interacting case under the same conditions. Other physical
parameters are identical to those in Fig. 7. Units are $J_0 =
0.773$~nA and $\Omega_0 = 77.3$~pW.}
\end{figure}

To further clarify the relationship between $\Omega_{op}$ and
$\Delta T$ shown in Fig.~8(f), Figs.~9(a)-9(c) present $S_{op} =
V_{op}/\Delta T$, $J_{op}$, and $\Omega_{op}$ as functions of
$\Delta T$ for $V_g = -32$, $-46$, and $-60$~mV. For $V_g =
-60$~mV, $\Omega_{op}$ reaches its maximum value at $k_B\Delta T =
20$~meV. When the temperature bias ($\Delta $T) approaches zero,
the Seebeck coefficient ($S_{op}=V_g / T_0$) (for $V_g = 60 $~mV)
depends only on the separation between the TDQD energy levels and
the Fermi energy ($\mu = 0$), as well as the average temperature
($T_0$). This indicates that the Seebeck coefficient of the TDQD
in the weak-coupling limit ($t_x = \Gamma_{et} \rightarrow 0$) can
function as an ultra-sensitive thermal detector
[\onlinecite{PreetiM}]. As shown in Fig.~9(a), $S_{op}$ is only
weakly dependent on $\Delta T$ for $V_g = -32$~mV, which also
explains why $V_{op}$ exhibits a linear dependence on $\Delta T$
for $V_g > -24$~mV in Fig.~8(d). When the TDQD energy levels lie
far from the Fermi energy ($V_g = -60$~mV), $S_{op}$ remains
linear in $\Delta T$, and this linearity persists up to $k_B\Delta
T = 20$~meV. It is noteworthy that $k_B\Delta T/(k_BT_0) \ge
\frac{10}{24}\ge\frac{5}{12}$ is not a perturbative parameter, yet
the nonlinear correction to $V_{op}$ is restricted to a quadratic
term, with no observable cubic or higher-order contributions which
is an unexpected behavior. This finding is consistent with the
recent analytical results reported in
Ref.~[\onlinecite{RaymondJH}], where a single noninteracting
quantum dot with one energy level was considered. In Fig.~9(b), we
find that $J_{op}$ remains a linear function of $\Delta T$ even at
$V_g = -60$~mV. Based on the trends identified in Fig.~9(a),
$\Omega_{op} = -J_{op}V_{op}$ for $V_g = -46$ and $-60$~mV, with
$\Omega_{op,\max}$ deviating from the expected $(\Delta T)^2$
scaling due to nonlinearities in $V_{op}$.

\begin{figure}[h]
\centering
\includegraphics[trim=1.cm 0cm 1.cm 0cm,clip,angle=0,scale=0.3]{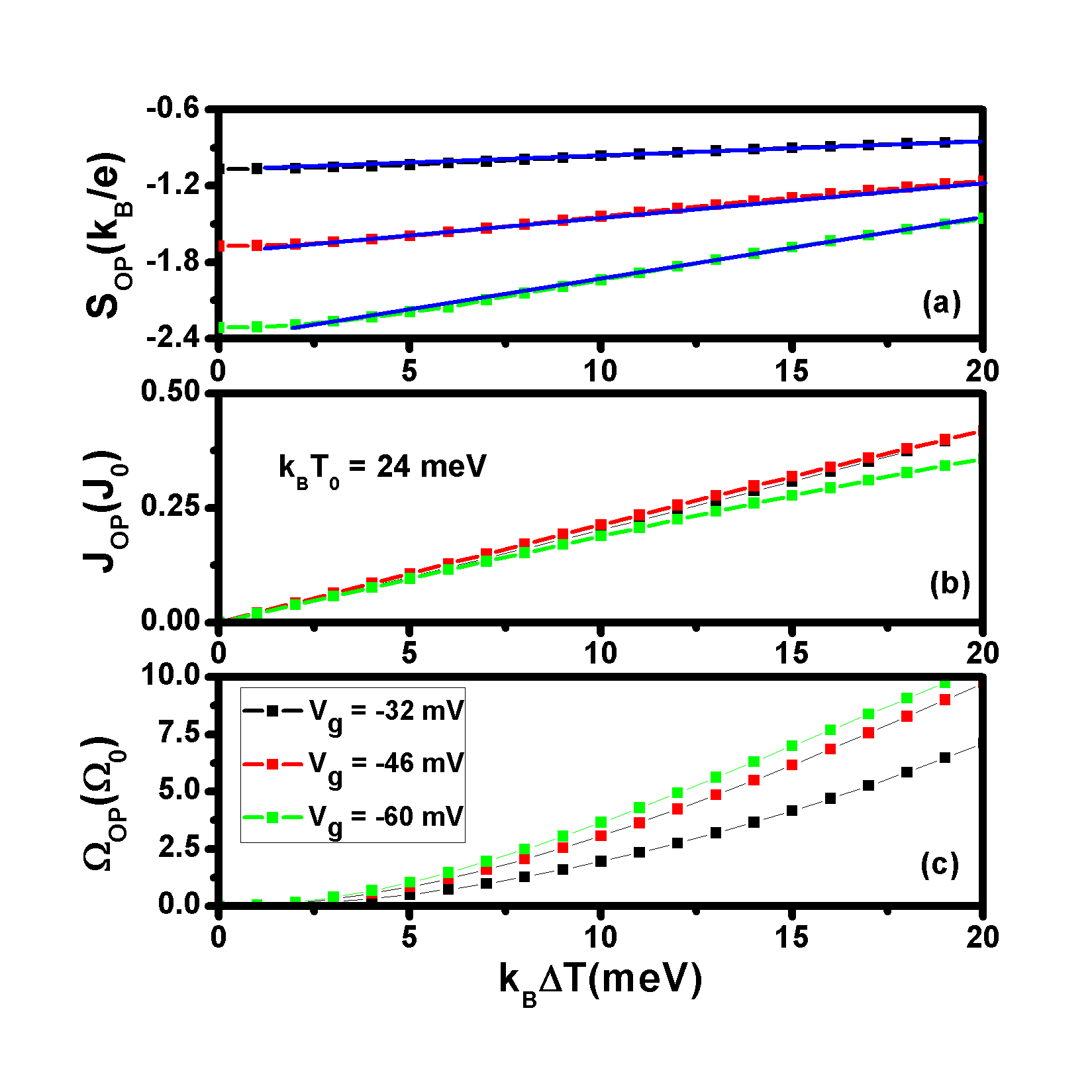}
\caption{(a) Optimized thermal voltage $S_{op} = V_{op}/\Delta T$,
(b) thermal current $J_{op}(\Delta T)$, and (c) electrical power
output $\Omega_{op}(\Delta T) = -J_{op}(\Delta T)\times
V_{op}(\Delta T)$ as functions of $\Delta T$ for $V_g = -32$,
$-36$, and $-60$~mV at $T_0 = 288$~K. Other physical parameters
are the same as in Fig.~8. The units are $J_0 = 0.773$~nA and
$\Omega_0 = 77.3$~pW.}
\end{figure}

\textbf{3. Electrical power rectification}\\

Finally, we examine the nonlinear thermoelectric properties of the
TDQD under asymmetric conditions, $V_{L,g} \neq V_{R,g}$ and
$\Gamma_{e,L} \neq \Gamma_{e,R}$. Figures 10(a)-10(d) show 2D
plots of total occupation number $N_t$, optimized thermal voltage
$V_{op}(\Delta T)$, thermal current $J_{op}(\Delta T)$, and
electrical power $\Omega_{op}(\Delta T)$ as functions of
temperature bias $\Delta T$ and interdot hopping $t_x$, for
parameters $\Gamma_{e,L} = t_x$, $\Gamma_{e,R} = t_x/10$,
$E_L=2$~meV, $E_R=8$~meV, and $T_0 = 48$~K. In Fig. 10(a), $N_t$
decreases with forward temperature bias ($\Delta T > 0$) and
increases with backward bias ($\Delta T < 0$) at weak hopping
($t_x = 1$~meV). $N_t$ is primarily determined by $N_L$, as the
left dot energy level is close to the Fermi energy ($\mu = 0$). At
stronger coupling ($t_x = 10$~meV), the broadening of $E_L$ due to
$\Gamma_{e,L} = t_x$ makes $N_L$ nearly independent of bias
direction, explaining the reduced sensitivity of $N_t$ to $\Delta
T$ sign.

The optimized thermal voltage $V_{op}$ is less sensitive to the
direction of $\Delta T$ (Fig. 10(b)). In contrast, $J_{op}$ and
$\Omega_{op}$ exhibit pronounced rectification (Figs. 10(c) and
10(d)), providing strong evidence that electron Coulomb
interactions govern power rectification; asymmetrical structures
alone cannot induce this effect in noninteracting nanostructures.
The asymmetry of $J_{op}$ and $\Omega_{op}$ can be understood from
the first term of the transmission coefficient:

\begin{eqnarray}
& &{\cal T}_{LR}(\varepsilon)/(4\Gamma_{e,L}t^2_x\Gamma_{e,R})\\
\nonumber &\approx&
\frac{1-N_{L,\sigma}-N_{R,\sigma}-N_{R,-\sigma}}{|(\varepsilon-E_L+i\Gamma_{e,L})(\varepsilon-E_R+i\Gamma_{e,R})-t^2_x|^2}
\end{eqnarray}
and
\begin{eqnarray}
& &{\cal T}_{RL}(\varepsilon)/(4\Gamma_{e,L}t^2_x\Gamma_{e,R})\\
\nonumber &\approx&
\frac{1-N_{R,\sigma}-N_{L,\sigma}-N_{L,-\sigma}}{|(\varepsilon-E_L+i\Gamma_{e,L})(\varepsilon-E_R+i\Gamma_{e,R})-t^2_x|^2}.
\end{eqnarray}
For $\Delta T > 0$, the relevant probability $P_1 \approx
1-N_{L,\sigma}-N_{R,\sigma}-N_{R,-\sigma}$, while for $\Delta T <
0$, $P_1 \approx 1-N_{R,\sigma}-N_{L,\sigma}-N_{L,-\sigma}$. Since
$N_L(\Delta T < 0) \gg N_L(\Delta T > 0)$, this accounts for the
observed asymmetry in $J_{op}$ and $\Omega_{op}$. Rectification of
$J_{op}(\Delta T)$ and $\Omega_{op}(\Delta T)$ requires TDQDs with
strong electron Coulomb interactions, asymmetric gate voltages,
and asymmetric 15-AGNR segment lengths.

\begin{figure}[h]
\centering
\includegraphics[angle=0,scale=0.3]{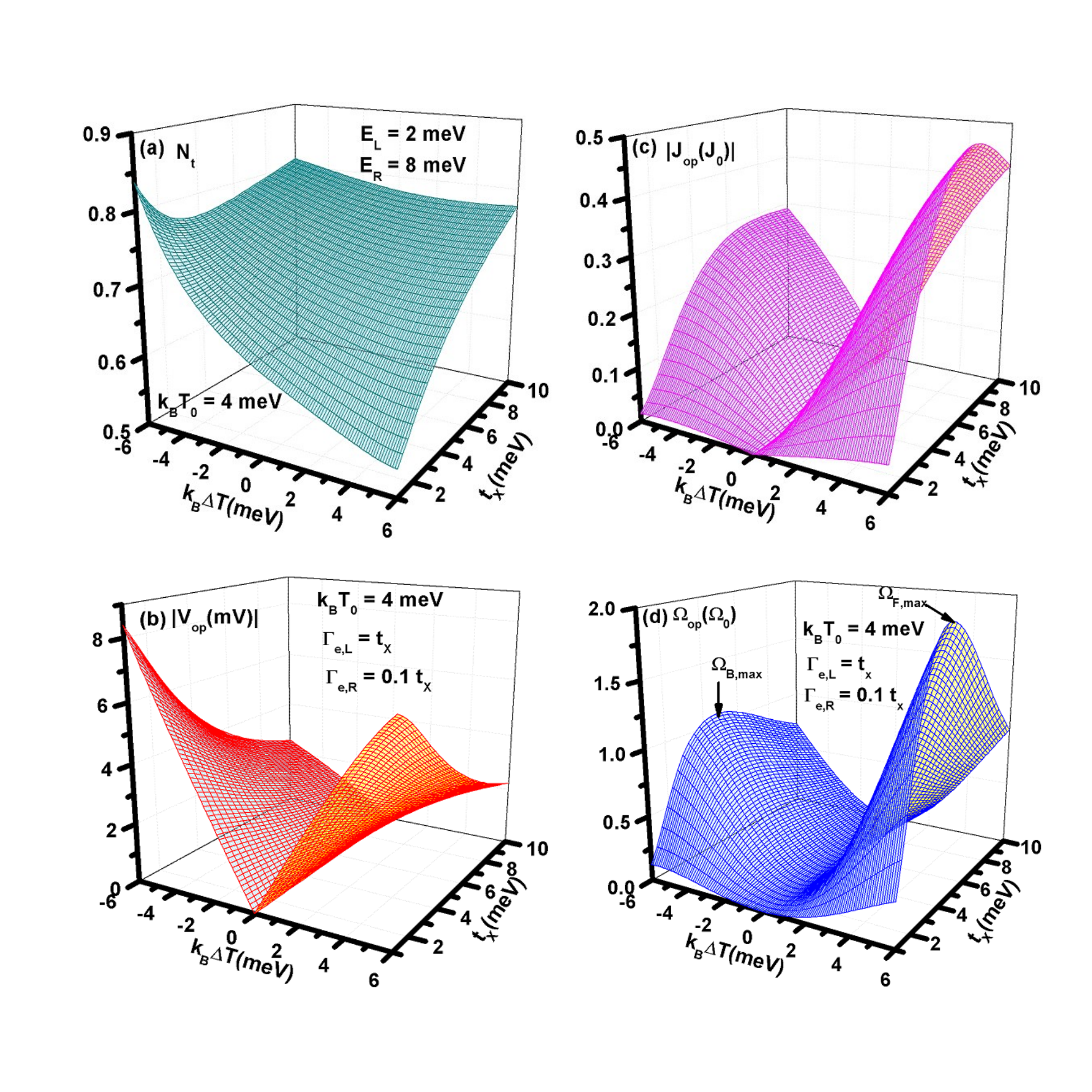}
\caption{(a) Total occupation number $N_t$, (b) optimized thermal
voltage $|V_{op}(\Delta T,T_0)|$, (c) optimized thermal current
$|J_{op}(\Delta T, T_0)|$, and (d) optimized power output
$\Omega_{op}(\Delta T, T_0) = -J_{op}(\Delta T, T_0)\times
V_{op}(\Delta T, T_0)$ as functions of temperature bias $\Delta T$
and interdot hopping $t_x$ at $k_B T_0 = 4$~meV, $E_L = 2$~meV,
$E_R = 8$~meV, $\Gamma_{e,L} = t_x$, and $\Gamma_{e,R} = 0.1~t_x$.
Other physical parameters are the same as in Fig. 8. Units are
$J_0 = 0.773$~nA and $\Omega_0 = 77.3$~pW.}
\end{figure}

\section{Conclusion}
In this study, we systematically clarified the number and types of
interface states (IFs) that emerge in N-AGNR/$(N-2)$-AGNR/N-AGNR
heterostructure (AGNRH) segments lacking translational symmetry,
based on their real-space geometric structure. We established a
direct correspondence between the end states (ESs) of individual
AGNR segments and the topological IFs of the resulting AGNRHs. For
AGNR segments with $R_1$-type unit cells, the numbers of ESs were
determined under a uniform longitudinal electric field for $N =
13, 15, 19, 21, 25,$ and $27$. These ESs satisfy the relations $N
= N_{A(B)} \times 6 + 1$ and $N = N_{A(B)} \times 6 + 3$, where
$N_{A(B)}$ denotes the number of ESs with A-(B-) chirality. For
AGNR segments with $R_2$-type unit cells, the total number of ESs
increases to $N_{A(B)} + 1$.

The Stark effect induced by the applied electric field lifts the
degeneracy of ESs in AGNRH segments and enables clear spectral
distinction between ESs and IFs. The real-space bulk boundary
perturbation approach further shows that states with opposite
chirality can hybridize through junction-site perturbations and
shift out of the bulk gap. We demonstrated that the number and
chirality of IFs, $N_{IF,\beta}$, in symmetric AGNRHs composed of
$N = 3p$ and $N = 3p+1$ AGNR segments are fully determined by the
ESs belonging to the outer and central AGNR components. Denoting
by $N_O$ and $N_C$ the numbers of ESs of the outer and central
AGNR segments, respectively, the IF number satisfies $N_{IF,\beta}
= |N_{O,B(A)}-N_{C,A(B)}|$, where $\beta$ specifies the chirality
of the IFs. When $N_{O,B} > N_{C,A}$, the IFs exhibit B-chirality,
while they acquire A-chirality when $N_{C,A} > N_{O,B}$. For
instance, the IFs of a 15-13-15 AGNRH segment originate from the
ESs of the central 13-AGNR segment with $R_2$-type unit cells,
whereas the IFs of a 27-13-27 AGNRH segment arise from the ESs of
the outer 27-AGNR segments.

Using the calculated transmission coefficient ${\cal
T}_{GNR}(\varepsilon)$ for
$N_{out}$-AGNR/$N_{cen}$-AGNR/$N_{out}$-AGNR segments with zigzag
edge structures coupled to electrodes, we demonstrated that IFs
act as a topological double quantum dot (TDQD) when the IFs are
formed by the ESs of the central AGNR segment. By employing an
Anderson model incorporating effective intradot and interdot
Coulomb interactions, we derived an analytical expression for the
tunneling current through the TDQD via the transmission
coefficient ${\cal T}_{LR}(\varepsilon)$. The thermoelectric
performance of AGNRH-based TDQDs was further analyzed in the
context of their potential application as graphene-nanoribbon
power generators~[\onlinecite{PerrinMM}].

The nonlinear thermoelectric power output of TDQDs exhibits
several notable features. (a) The optimized power output is
favored in either the electron-dilute $(0,0)$ or hole-dilute
$(2,2)$ charge configurations. In the $(0,0)$ configuration,
Coulomb blockade strongly suppresses $J_{op,\max}$ while leaving
$V_{op,\max}$ largely unaffected. (b) Nonlinear temperature bias
induces only a quadratic correction to the thermal voltage
$V_{op}$, with no cubic or higher-order terms observed. (c) Even
in the presence of strong electron electron Coulomb interactions,
the thermal power output remains highly enhanced under nonlinear
temperature bias. (d) Direction-dependent power output arises from
strong electron correlations and the asymmetry in outer AGNR
lengths. Owing to their well-isolated energy levels located deep
within the bulk gap of 15-13-15 AGNRHs, TDQDs represent promising
candidates for high-temperature thermoelectric power generation.

\textbf{Data availability}\\
The data that supports the finding of this study are available
within the article

\textbf{Conflicts of interest}\\
There are no conflicts to declare.

{\bf Acknowledgments}\\
This work was supported by the National Science and Technology
Council, Taiwan under Contract No. MOST 107-2112-M-008-023MY2.

{}
\mbox{}\\


\appendix
\numberwithin{figure}{section}

\numberwithin{equation}{section}


\section{Interface States of Wide-AGNR/Narrow-AGNR/Wide-AGNR AGNRH Segments}

For N-AGNR/$(N-2)$-AGNR/N-AGNR AGNRH segments, the width
difference $\Delta N = (N_{out}-N_{cen})/2$ equals $\Delta N = 1$,
where $N_{out}$ and $N_{cen}$ denote the atom widths of the outer
and central AGNR segments, respectively. In Fig.~4, our analysis
focused on the case $\Delta N = 1$. In this appendix, we examine
how increasing $\Delta N$ affects both the number and chirality of
interface states (IFs) in AGNRHs. Figures~A.1(a-d) display the
energy spectra of four AGNRH segments as functions of $t_{es}$:
(a) $27_4-25_{10}-27_4$, (b) $27_4-21_{10}-27_4$, (c)
$27_4-13_{10}-27_4$, and (d) $27_4-9_{10}-27_4$. These correspond
to $\Delta N = 1, 3, 7$, and $9$, respectively, with a fixed outer
width of $N_{out} = 27$. In all cases, the outer AGNRs have an
$R_1$-type unit cell, whereas the central AGNR adopts an
$R_2$-type unit cell.

In Fig.~A.1(a), one interface state of sublattice-A (sublattice-B)
arises at the 27-AGNR/25-AGNR (25-AGNR/27-AGNR) junction. The
resulting pair of IF levels, $\Sigma_{IF,c}$ and $\Sigma_{IF,v}$,
originates from their mutual coupling. Compared with the
$27_4-25_6-27_4$ case in Fig.~4(d), the energy separation between
$\Sigma_{IF,c}$ and $\Sigma_{IF,v}$ in Fig.~A.1(a) is smaller due
to the longer $25_{10}$ central segment. In Fig.~A.1(b), eight
in-gap states appear, all attributable to the end states of the
$27-21-27$ AGNRH system. Since both 27-AGNR and 21-AGNR segments
have $N_{A(B)}=4$, the number of IFs is zero, consistent with the
rule that IFs are determined by the difference in ES counts. In
Fig.~A.1(c), a single interface state with sublattice-B(A)
character is present at each 27-AGNR/13-AGNR (13-AGNR/27-AGNR)
junction because the 13-AGNR segment has $N_{A(B)} = 3$. These IFs
originate from the ESs of the outer AGNRs; consequently, their
mutual overlap is extremely small, whereas the IFs strongly couple
to the terminal states of the 27-13-27 AGNRH segment. Their
evolution with respect to $t_{es}$ is therefore markedly different
from IFs originating from the central AGNR segment. In
Fig.~A.1(d), two IFs appear at the 27-AGNR/9-AGNR (9-AGNR/27-AGNR)
junctions because the 9-AGNR segment possesses $N_{A(B)} = 2$.

\begin{figure}[h]
\centering
\includegraphics[angle=0,scale=0.3]{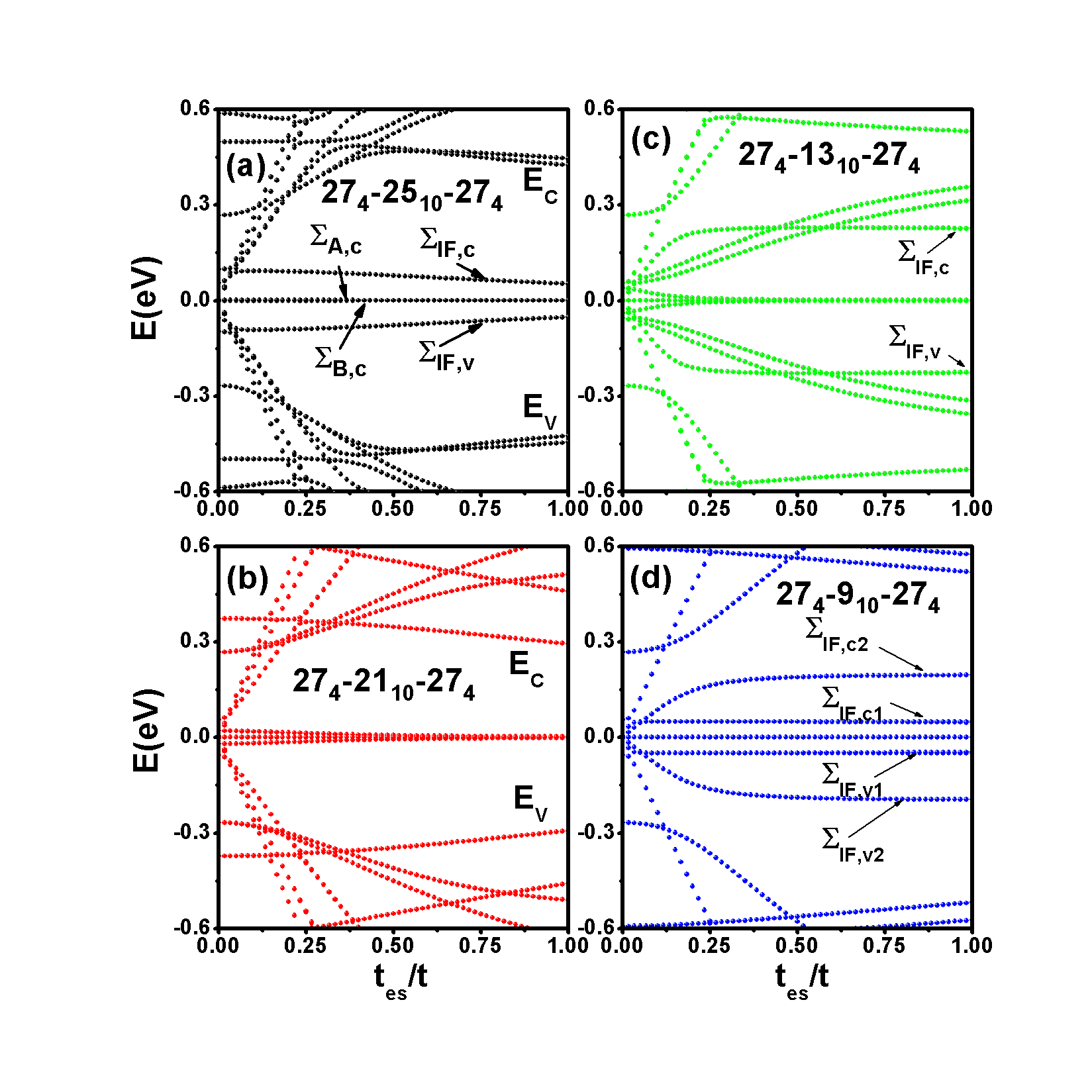}
\caption{Energy levels of four AGNRH segments with varying central
widths as functions of the inter-AGNR electron hopping strength
$t_{es}$: (a) $27_4-25_{10}-27_4$, (b) $27_4-21_{10}-27_4$, (c)
$27_4-13_{10}-27_4$, and (d) $27_4-9_{10}-27_4$ AGNRH segments.}
\end{figure}

A particularly large energy separation between $\Sigma_{IF,c2}$
and $\Sigma_{IF,v2}$ is seen in Fig.~A.1(d). To understand its
origin, Fig.~A.2 presents the energy levels of $27_w-9_6-27_w$
AGNRHs for four different outer lengths $w = 6, 7, 8$, and $9$. As
$w$ increases, the splitting between $\Sigma_{IF,c2}$ and
$\Sigma_{IF,v2}$ decreases significantly. This trend indicates
that these two levels arise from IFs coupled to end states of the
27-AGNR segments that have long decay lengths. In contrast, the
splitting between $\Sigma_{IF,c1}$ and $\Sigma_{IF,v1}$ shows only
weak dependence on $w$, implying that these states originate from
IFs associated with short-decay-length end states of the 27-AGNR
segments.

\begin{figure}[h]
\centering
\includegraphics[angle=0,scale=0.3]{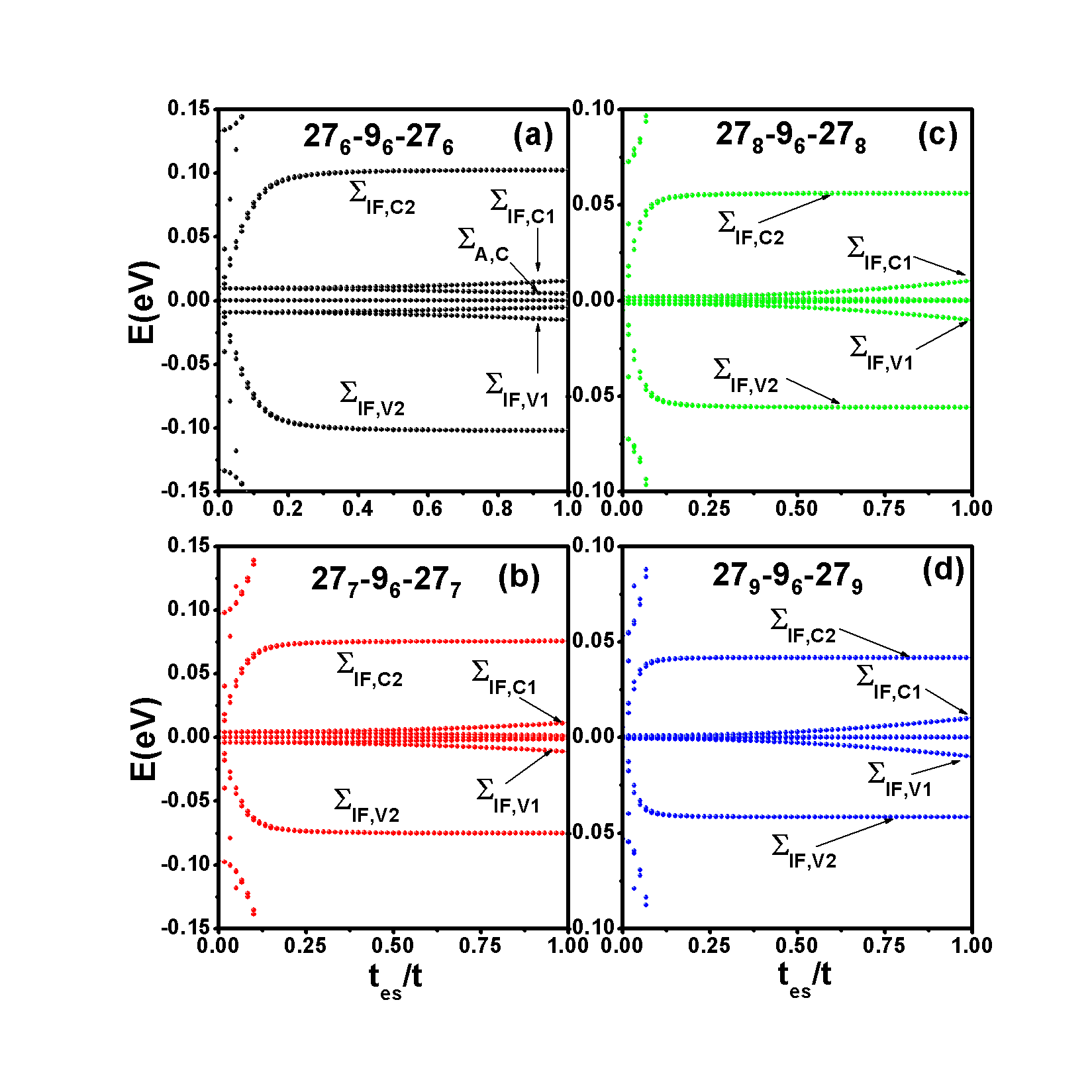}
\caption{Energy levels of $27_w-9_6-27_w$ AGNRH segments as
functions of the inter-AGNR hopping strength $t_{es}$ for
different outer 27-AGNR lengths: (a) $27_6-9_6-27_6$, (b)
$27_7-9_6-27_7$, (c) $27_8-9_6-27_8$, and (d) $27_9-9_6-27_9$
AGNRH segments.}
\end{figure}

Figure~A.3 shows the probability densities of representative IF
levels specifically, the pair $\Sigma_{IF,c2,a} = 102.2335$~meV
and $\Sigma_{IF,c2,b} = 102.2232$~meV, the level $\Sigma_{IF,c1} =
15.4212$~meV, and the terminal-state level $\Sigma_{A,c} =
0.539$~meV corresponding to Fig.~A.2(a). In Figs.~A.3(a) and
A.3(b), the spatial distributions of $\Sigma_{IF,c2,a}$ and
$\Sigma_{IF,c2,b}$ are nearly indistinguishable; furthermore,
their amplitudes inside the central 9-AGNR segment are extremely
weak, indicating that transport through $\Sigma_{IF,c2}$ is
strongly suppressed. In contrast, $\Sigma_{IF,c1}$ exhibits a
pronounced probability density within the central 9-AGNR segment,
making it a more favorable transport channel. As shown in
Fig.~A.3(c), $\Sigma_{IF,c1}$ effectively behaves as a set of four
topological dots (TDs). Figure~A.3(d) demonstrates that the state
$\Sigma_{A,c}$ is primarily associated with the terminal states of
the 27-9-27 AGNRH structure.

\begin{figure}[h]
\centering
\includegraphics[angle=0,scale=0.25]{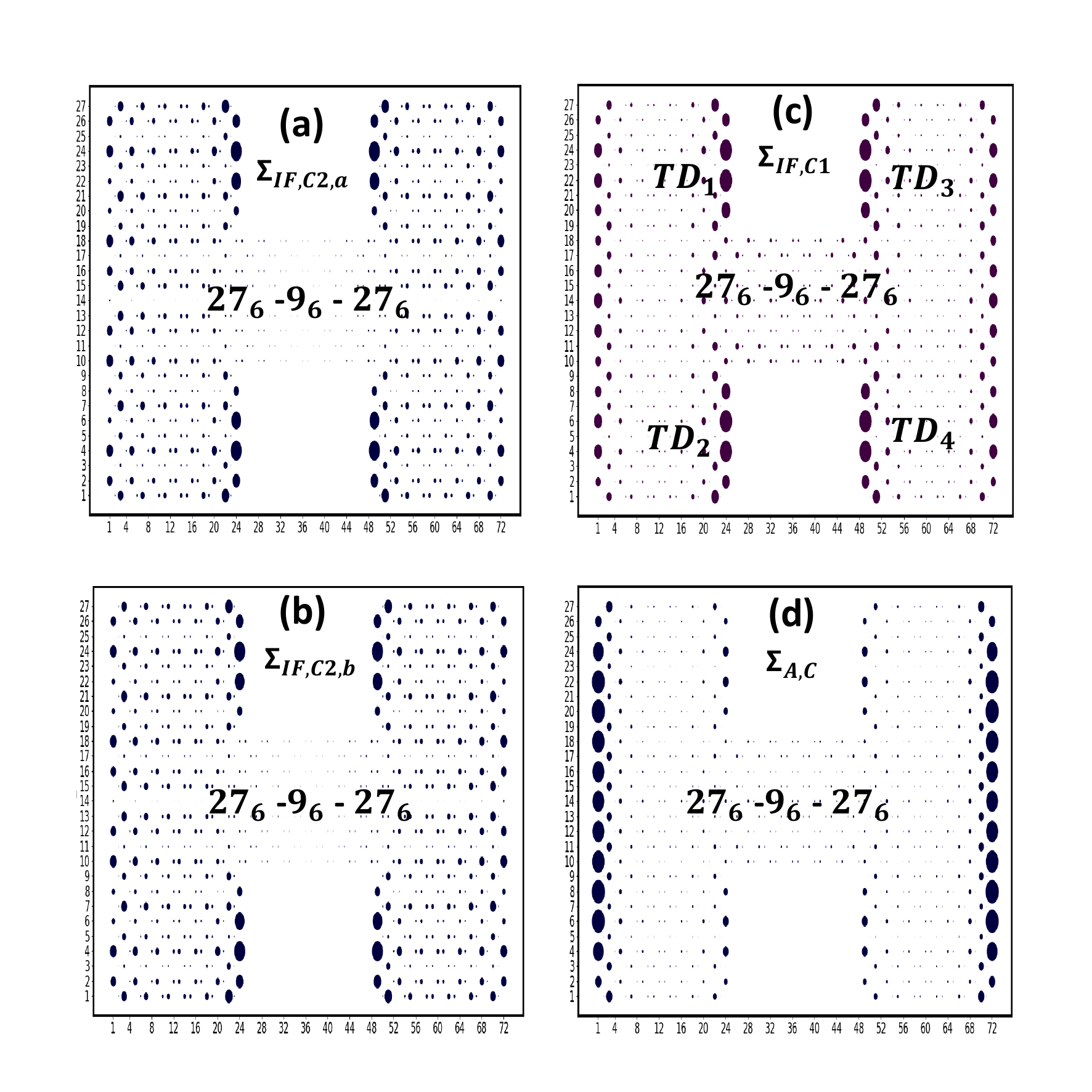}
\caption{ Probability densities of selected energy levels in the
$27_6-9_6-27_6$ AGNRH segment with $t_{es} = t$: (a)
$\Sigma_{IF,c2,a} = 102.2335$~meV, (b) $\Sigma_{IF,c2,b} =
102.2232$~meV, (c) $\Sigma_{IF,c1} = 15.4212$~meV, and (d)
$\Sigma_{A,c} = 0.539$~meV.}
\end{figure}





\end{document}